\documentclass {tlp}
\usepackage {graphicx}
\usepackage {harvard}

\newtheorem{definition}{Definition}
\newtheorem{example}{Example}

\begin{document}
\title[On the Abductive or Deductive Nature of Database Problems]
{On the Abductive or Deductive Nature of Database Schema Validation and Update Processing Problems}

\author[E.Teniente and T.Urp\'{\i}]
{Ernest Teniente and Toni Urp\'{\i}\\
Departament de Llenguatges i Sistemes Inform\`{a}tics \\
Universitat Polit\`{e}cnica de Catalunya\\
Jordi Girona Salgado 1-3, Barcelona, Catalonia\\ 
e-mail [teniente $\vert $ urpi]@lsi.upc.es\\
}

\pagerange{\pageref{firstpage}--\pageref{lastpage}}
\volume{\textbf{10} (3):}
\jdate{March 2000}
\setcounter{page}{1}
\pubyear{2000}

\maketitle
\label{firstpage}

\begin{abstract}
We show that database schema validation and update processing problems such as view updating, materialized view maintenance, integrity constraint checking, integrity constraint maintenance or condition monitoring can be classified as problems of either abductive or deductive nature, according to the reasoning paradigm that inherently suites them. This is done by performing abductive and deductive reasoning on the event rules \cite{Oli91}, a set of rules that define the difference between consecutive database states In this way, we show that it is possible to provide methods able to deal with all these problems as a whole.

We also show how some existing general deductive and abductive procedures may be used to reason on the event rules. In this way, we show that these procedures can deal with all database schema validation and update processing problems considered in this paper.
\end{abstract}

\section{Introduction}

It is largely agreed that databases should contain, at least, base facts, 
deductive rules (views), integrity constraints and, sometimes, conditions to 
monitor since these features, together with appropriate reasoning 
capabilities, facilitate program development and reuse \cite{GM92}. Base facts 
represent extensional information; while deductive rules, integrity 
constraints and conditions to monitor represent intentional information, 
i.e. information that can be inferred from the extensional one. In 
particular, deductive rules define common knowledge shared by different 
users; integrity constraints define conditions that each state of the 
database is required to satisfy and conditions to monitor define information 
whose changes must be notified to the user.

Database schema validation has become an important problem in database 
engineering, particularly since databases are able to define intentional 
information. Schema validation refers to the process of verifying whether a 
database schema correctly and adequately describes the user's intended needs 
and requirements \cite{ABC82}. In general, preventing possible flaws during 
schema design will prevent those flaws from materializing as execution time 
errors or inconveniences. Among the typical problems related to database 
schema validation, we have satisfiability checking, redundancy of integrity 
constraints, view liveliness, etc. \cite{BM86,LS95,DTU96}.

Databases must also include an update processing system that provides users 
with a uniform interface through which they can request different kinds of 
updates, e.g. updates of base facts or updates of derived facts. In the 
presence of intentional information, updating a database is not a simple 
task since many issues have to be taken into account \cite{Abi88}. Therefore, an 
important amount of research has been devoted to different database updating 
problems like view updating \cite{KM90,GL91,TO95,CST95,Dec96,LT97}, 
materialized view maintenance \cite{GM95,Ros98}, integrity constraint checking 
\cite{SK88,Kuc91,Oli91,GCM94,LL96,SJ96}, integrity constraint 
maintenance \cite{ML91,CFPT94,Wut93,ST94,TO95} or condition monitoring 
\cite{RCB89,HCK90,QW91,BCP98}. 

Up to now, the general approach of the research related to database schema 
validation and update processing has been to provide specific methods for 
solving particular problems. Therefore, if we were interested in 
integrating these problems into current database technology, we should 
incorporate several methods into commercial database management systems. In 
our opinion, this is one of the main reasons of the difficulty of 
translating the huge amount of research in this area into practical 
applications. 

Solving these problems requires reasoning about the effect of an update on 
the database. Therefore, all these methods are either explicitly or 
implicitly based on a set of rules that define the changes that occur in a 
transition from an old state of a database to a new one, which is obtained 
as a result of the application of a certain transaction consisting of a set 
of base fact updates. 

Several authors have argued about the advantages of making explicit the 
rules that define changes on the database contents induced by the 
application of a transaction when dealing with database updating problems 
\cite{Bry90,TU95,DD98}. These rules allow to guarantee that the updated 
database is as close as possible to the old database (which is the 
traditional assumption considered in database updating) and provide a high 
level of expressiveness since they allow to reason jointly about both states 
involved in the update (which is specially useful when dealing with database 
updating problems).

On the other hand, the role of deduction and abduction as reasoning 
paradigms is widely accepted. For instance, deduction has been used for 
query processing, while abduction has been applied to fault diagnosis, 
planning or default reasoning. In the context of databases, abductive 
procedures have been proposed for view updating \cite{Bry90,CST95,Dec96} or 
satisfiability checking \cite{DD98}. However, we do not know precisely how many 
forms of reasoning are necessary for solving known database problems nor, in 
general, which database problems can be considered of deductive or of 
abductive nature, according to the reasoning paradigm that is more naturally 
suited to solve them.

In this paper we show that database schema validation and update 
processing problems can be classified as either of 
deductive or of abductive nature. This is done by considering the event 
rules \cite{Oli91}, a particular case of rules that define the exact difference 
among consecutive database states, and by showing that reasoning deductively 
or abductively on these rules allows us to naturally specify and handle 
database problems. As we will see, problems like materialized view 
maintenance, integrity constraint checking or condition monitoring will be 
considered as naturally deductive, while problems like view updating, 
integrity constraint maintenance or enforcing condition activation as 
naturally abductive.

This first result is an evolution of our work in \cite{TU95} where we proposed 
two ad-hoc procedures to reason about the event rules that allowed us to 
classify database updating problems. We extend this work by showing that, in 
fact, we do not need ad-hoc procedures but that we can consider general 
reasoning paradigms like deduction and abduction. It follows from this 
result that general deductive and abductive procedures can be used to reason 
about the event rules and, hence, to deal with database schema validation 
and update processing problems.

We also show how some existing general deductive and abductive procedures 
may be used to reason on the event rules. In this way, we show that these 
procedures can be used to deal with all database schema validation and 
update processing problems considered in this paper. This is illustrated by 
means of examples and we also point out some additional benefits gained by 
these procedures when reasoning on the event rules. Note that our goal is 
not that of comparing existing procedures but to show how to use them to 
reason on the event rules.

Finally, we sketch how the event rules could be used to solve general 
abductive problems in addition to database schema validation and update 
processing problems. 

Problems related to views, integrity constraints and conditions to be 
monitored will be encountered whenever we have a database able to deal with 
intentional information. We have developed our ideas for the particular case 
of deductive databases due to their clear and precise notation. However, our 
framework (and, thus, our conclusions) can be easily generalized to all 
kinds of databases containing views, integrity constraints and conditions 
like, for instance, relational, active, object or object-relational 
databases.

This paper is organized as follows. The next section reviews basic concepts 
of deductive databases. Section 3 shortly presents the concepts of event, 
transition rules and event rules. Section 4 defines deductive and abductive 
reasoning on the event rules. Section 5 describes the most important 
problems related to schema validation and update processing and
 classifies them as either of 
deductive or of abductive nature. Section 6 shows how to use general 
deductive and abductive procedures to reason on the event rules. Section 7 
sketches the use of the event rules to solve general abductive problems. 
Finally, in section 8 we present our conclusions and point out future work.

\section{Basic Definitions and Notation}

We briefly review the basic concepts of deductive databases \cite{Llo87,ull89}. 
We consider a first order language with a universe of constants, a set of 
variables, a set of predicate names and no function symbols. We will use 
names beginning with a capital letter for predicate symbols and constants 
and names beginning with a lower case letter for variables.

A \textit{term} is a variable symbol or a constant symbol. We assume that possible values 
for terms range over finite domains. If P is an m-ary predicate symbol and 
$t_{1}, \ldots, t_{m}$ are terms, then $P(t_{1}, \ldots, t_{m})$ is an \textit{atom}. The 
atom is \textit{ground} if every $t_{i}$ (i = 1, $\ldots$ , m) is a constant. A \textit{literal} is defined 
as either an atom or a negated atom.

A \textit{fact} is a formula of the form: P(t$_{1}$, $\ldots$, t$_{m})$ $\leftarrow$, 
where P(t$_{1}$, $\ldots$, t$_{m})$ is a ground atom. We will omit the arrow when 
denoting an atom.

A \textit{deductive rule} is a formula of the form: $P(t_{1},\ldots, t_{m})
\leftarrow L_{1} \wedge \ldots \wedge L_{n},$ with m $ 
\ge $ 0, n $ \ge $ 1, where $P(t_{1}, \ldots, t_{m})$ is an atom denoting 
the \textit{conclusion} and L$_{1}, \ldots, L_{n}$ are literals representing 
\textit{conditions}. P(t$_{1}, \ldots, t_{m})$ is called the \textit{head} 
and L$_{1} \wedge \ldots \wedge L_{n}$ the \textit{body} of the 
deductive rule. Variables 
in the conclusion or in the conditions are assumed to be universally 
quantified over the whole formula. The definition of a predicate P is the 
set of all rules in the database which have P in their head. We assume that 
the terms in the head are distinct variables. 

An \textit{integrity constraint} is a formula that every state of the database is required to satisfy. We 
deal with constraints in \textit{denial} form: $ \leftarrow L_{1} \wedge \ldots \wedge L_{n}$, with n $\ge$ 1, 
where the L$_{i}$ are literals and all variables are assumed to be 
universally quantified over the whole formula. We associate to each 
integrity constraint an inconsistency predicate Icn, where n is a positive 
integer, and thus they have the same form as deductive rules. Then, we would 
rewrite the former denial as Ic1 $ \leftarrow L_{1} \wedge \ldots \wedge L_{n}$. We call 
them \textit{integrity rules}. More general constraints can be transformed into denial form by 
applying the procedure described in \cite{LT84}.

We also assume that the database contains a distinguished derived predicate 
Ic defined by the n clauses: Ic $ \leftarrow $ Icn. That is, one rule for 
each integrity constraint Ici, i=1...n, of the database. Note that Ic will 
only hold in those states of the database that violate some integrity 
constraint and that it will not hold for those states that satisfy all 
constraints.

A \textit{condition to be monitored} is a formula of the form: $Cond(t_{1}, ..., t_{m})
\leftarrow L_{1} \wedge \ldots \wedge L_{n}$, with m $ 
\ge $ 0, n $ \ge $ 1, where $Cond(t_{1}, ..., t_{m})$ is an atom and 
L$_{1}, \ldots, L_{n}$ are literals. Moreover, variables in 
Cond and in L$_{1}, \ldots, L_{n}$ are assumed to be universally quantified over the whole formula. 
Each condition to be monitored corresponds to a derived predicate for which 
certain changes have to be notified to the database user. 

A \textit{deductive database }D is a tuple D = (EDB, IDB) where EDB is a set of facts, and IDB = DR $ 
\cup $ IC $ \cup $ Cond is a set of rules such that DR is a set of deductive 
rules, IC a set of integrity rules and Cond a set of conditions to be 
monitored. The set EDB of facts is called the \textit{extensional} part of the deductive 
database and the set IDB of rules is called the \textit{intentional} part. We say that a 
deductive database is consistent if predicate Ic does not hold on it, i.e. 
if no integrity constraint is violated.

We assume that deductive database predicates are partitioned into base and 
derived (view) predicates. A base predicate appears only in the extensional 
part and (possibly) in the body of deductive rules. A derived predicate 
appears only in the intentional part. Base and derived facts correspond to 
facts about base and derived predicates, respectively.

As usual, we require that the deductive database before and after any 
updates is \textit{allowed}, that is, any variable that occurs in a deductive rule, 
integrity rule or condition to be monitored has an occurrence in a positive 
literal of the body of the rule. In this paper, we deal with 
hierarchical\footnote{ Equivalent to nr-datalog-$\neg $ rules, and with the 
same expressive power as the relational calculus \cite{AHV95}.} databases. Note 
that, in particular, this kind of databases does not allow to express 
recursive rules.

\section{The Event Rules}
\label{sectEventRules}

Intuitively, a database is a dynamic system that changes over time. Changes 
are effected by EDB updates. These updates define a transition from an old 
state of the database to a new updated one. In this sense, schema validation 
and update processing problems can be viewed as database state transition 
problems. It is possible to define a set of rules that explicitly defines 
the possible transitions in terms of the difference between consecutive 
database states. Reasoning about these rules will allow to reason jointly 
about both states involved in the transition and, thus, to reason about the 
effect of the updates.

The event rules explicitly define the difference between two consecutive 
database states. In the following, we shortly review the concepts and 
terminology of event rules, as presented in \cite{Oli91}, and we discuss on the 
possible use of other sets of rules instead of the event rules.

\subsection{Events}

Let D$^{o}$ be a deductive database, T a transaction and D$^{n}$ the updated 
deductive database. We say that T induces a transition from D$^{o}$ (the old 
state) to D$^{n}$ (the new state). We assume for the moment that T consists 
of an unspecified set of base facts to be inserted and/or deleted.

Due to the deductive rules, T may induce other updates on some derived 
predicates. Let P be one of such predicates, and let P$^{o}$ and P$^{n}$ 
denote the same predicate evaluated in D$^{o }$ and $^{ }$D$^{n}$, 
respectively. Assuming that a fact P$^{o}$(\textbf{C}) holds in D$^{o}$, 
where \textbf{C} is a vector of constants, two cases are possible:

\begin{tabbing}
XXXXX\=\kill
\>a.1. P$^{n}$(\textbf{C}) also holds in D$^{n}$.\\
\>a.2. P$^{n}$(\textbf{C}) does not hold in D$^{n}$.
\end{tabbing}

\noindent
and assuming that P$^{n}$(\textbf{C}) holds in D$^{n}$, two cases are also 
possible:

\begin{tabbing}
XXXXX\=\kill
\>b.1. P$^{o}$(\textbf{C}) also holds in D$^{o}$. \\
\>b.2. P$^{o}$(\textbf{C}) does not hold in D$^{o}$.
\end{tabbing}

In case a.2 we say that a deletion event occurs in the transition, and we 
denote it by $\delta $P(\textbf{C}). In case b.2 we say that an insertion 
event occurs in the transition, and we denote it by $\iota $P(\textbf{C}). 

Formally, we associate to each predicate P an \textit{insertion event predicate} \textit{$\iota $P} and a \textit{deletion event predicate $\delta $P}, defined as:

\begin{tabbing}
XXXXX\=\kill
\>(1) ${\forall }$\textbf{x }($\iota $P(\textbf{x}) $ \leftrightarrow {\rm 
P}^{n}$(\textbf{x}) $ \wedge \: {\rm \neg}$P$^{o}$(\textbf{x})) \\

\>(2) ${\forall }$\textbf{x }($\delta $P(\textbf{x}) $ \leftrightarrow {\rm 
P}^{o}$(\textbf{x}) $ \wedge  \: {\rm \neg}$P$^{n}$(\textbf{x})) 
\end{tabbing}

\noindent
where \textbf{x} is a vector of variables. From the above, we then have the 
equivalencies:

\begin{tabbing}
XXXXX\=\kill
\>(3) ${\forall}$\textbf{x }(P$^{n}$(\textbf{x}) $ \leftrightarrow {\rm 
}$(P$^{o}$(\textbf{x}) $ \wedge \: {\rm \neg}\delta $P(\textbf{x})) $ \vee $ 
$\iota $P(\textbf{x})) \\

\>(4) ${\forall}$\textbf{x }($\neg $P$^{n}$(\textbf{x}) $ \leftrightarrow {\rm 
}(\neg $P$^{o}$(\textbf{x}) $ \wedge \: {\rm \neg}\iota $P(\textbf{x})) $ 
\vee \: \delta $P(\textbf{x})) 
\end{tabbing}

We say that an event $\iota $P or $\delta $P is a base event if P 
is a base predicate. Otherwise, it is a derived event. If P is a base 
predicate, then $\iota $P and $\delta $P facts represent insertions 
and deletions of base facts, respectively. Therefore, we assume from now on 
that a transaction T consists of an unspecified set of insertion and/or 
deletion base event facts. If P is a derived predicate, an integrity 
constraint or a condition to be monitored, $\iota $P and $\delta $P facts 
represent induced insertions and induced deletions on P, respectively.

\subsection{Transition Rules}

Let us consider a derived predicate P of the database. The definition of P 
consists of the rules in the deductive database having P in the conclusion. 
Assume that there are m (m $ \ge $ 1) such rules. For notation's sake, we 
rename predicate symbols in the conclusions of the m rules by P$_{1}$, ..., 
P$_{m}$, replace the implication by an equivalence and add the set of rules:

\begin{tabbing}
XXXXX\=XXXXXXXXXX\=\kill
\>(5) P $\leftarrow  P_{i} $ \> $i=1\ldots m$
\end{tabbing}
\noindent
i.e. one rule defining P for each derived predicate P$_{i}, i = 1\ldots m$.

Consider now one of the rules P$_{i}(\textbf{x}) \leftrightarrow  L_{1} \wedge \ldots \wedge L_{n}$.
When this rule is to be evaluated in the new state, its form is 
P$^{n}_{i}(\textbf{x}) \leftrightarrow L^{n}_{1} \wedge \ldots \wedge L^{n}_{n}$, where 
L$^{n}_{j}$ (j = 1 \ldots n) is obtained by replacing the predicate Q of 
L$_{j}$ by Q$^{n}$. Then, if we replace each literal in the body by its 
equivalent expression given in (3) or (4) we get a new rule which defines 
the new state predicate P$^{n}_{i}$ in terms of old state predicates and 
events. 

More precisely, if L$^{n}_{j }$ is a positive literal 
Q$^{n}_{j}$(\textbf{x}$_{j})$ we apply (3) and replace it by: 

\begin{tabbing}
XXXXX\=\kill
\>(Q$^{o}_{j}$(\textbf{x}$_{j}) \wedge \neg  \: \delta 
$Q$_{j}$(\textbf{x}$_{j}))$ $ \vee \: \iota $Q$_{j}$(\textbf{x}$_{j})$
\end{tabbing}
\noindent
and if L$^{n}_{j }$ is a negative literal $\neg 
$Q$^{n}_{j}$(\textbf{x}$_{j})$ we apply (4) and replace it by:

\begin{tabbing}
XXXXX\=\kill
\>($\neg $Q$^{o}_{j}$(\textbf{x}$_{j}) \wedge \neg \iota 
$Q$_{j}$(\textbf{x}$_{j}))$ $ \vee  \: \delta $Q$_{j}$(\textbf{x}$_{j})$ 
\end{tabbing}

After distributing $ \wedge $ over $ \vee $, we get the set of \textit{transition rules} for 
P$^{n}_{i}$. Notice that there are 2$^{ki}$ such rules (where k$_{i }$ is 
the number of literals in the P$^{n}_{i}$ rule) and that literals in each 
rule can be of three types: old database literals, base event literals and 
derived event literals.

\begin{example}
Consider the rule Cont$_{1}$(x) $ \leftrightarrow $ 
Sign(x) $ \wedge \neg$Fail-ex(x) stating that contracted people are 
those who signed an agreement and did not failed the exam. In the new state, 
this rule has the form Cont$^{n}_{1}$(x) $ \leftrightarrow $ Sign$^{n}$(x) 
$ \wedge \neg$Fail-ex$^{n}$(x). Then, replacing Sign$^{n}$(x) and $\neg 
$Fail-ex$^{n}$(x) by their equivalent expressions given by (3) and (4) we 
get:

\begin{tabbing}
XXXXX\=XXXXXXXx\=\kill
\>$Cont^{n}_{1}(x) \leftrightarrow [(Sign^{o}(x) \wedge \neg 
\delta Sign(x)) \vee \iota Sign(x)] \: \wedge$ \\

\>\>$[(\neg Fail$-$ex^{o}(x) \wedge \neg \iota Fail$-$ex(x)) \vee  
\delta Fail$-$ex(x)]$
\end{tabbing}

\noindent
and, after distributing $ \wedge $ over $ \vee $, we get the following 
transition rules:

\begin{tabbing}
XXXXX\=\kill
\>Cont$^{n}_{1}(x) \leftarrow Sign^{o}(x) \wedge \neg\delta Sign(x)
 \wedge \neg Fail$-$ex^{o}(x) \wedge \neg\iota Fail$-$ex(x) $\\
\>Cont$^{n}_{1}(x) \leftarrow Sign^{o}(x) \wedge \neg \delta Sign(x) \wedge \delta Fail$-$ex(x) $\\
\>Cont$^{n}_{1}(x) \leftarrow \iota Sign(x) \wedge \neg Fail$-$ex^{o}(x) \wedge \neg \iota Fail$-$ex(x) $\\
\>Cont$^{n}_{1}(x) \leftarrow \iota Sign(x) \wedge \delta $Fail-ex(x)
\end{tabbing}

Intuitively, it is not difficult to see that the first rule states that 
Cont(X) will be true in the new state of the database if Sign(X) was true in 
the old state, Fail-ex(X) was false in the old state and no change of 
Sign(X) and Fail-ex(X) is given by the transition. In a similar way, the 
second rule states that Cont(X) will be true in the new state if Sign(X) was 
true and it has not been deleted and if Fail-ex(X) has been deleted during 
the transition. A similar, intuitive, interpretation can be given for the 
third and forth rules.
\end{example}

For simplicity of presentation, we will omit the subscript when a predicate 
P is defined by only one rule and we will omit the superscript$^{o}$ for 
denoting old database predicates.

\subsection{Insertion and Deletion Event Rules}

Let P be a derived predicate. Insertion and deletion event rules of 
predicate P are defined respectively as:

\begin{tabbing}
XXXXX\=\kill
\>(6) $\iota $P(\textbf{x}) $ \leftarrow {\rm 
P}^{n}$(\textbf{x}) $ \wedge  {\rm \neg}$P$^{o}$(\textbf{x}) \\

\>(7) $\delta $P(\textbf{x}) $ \leftarrow {\rm 
P}^{o}$(\textbf{x}) $ \wedge  {\rm \neg}$P$^{n}$(\textbf{x}) 
\end{tabbing}

\noindent
where P$^{ }$ refers to the current (old) database state and P$^{n}$ refers 
to the transition rule of P. These event rules define the induced changes 
that happen in a transition from an old state of a database to a new, 
updated state. Note that they depend only on the rules of the database, 
being independent of the stored facts and of any particular transaction.

We would like to point out that these rules can be intensively simplified, 
as described in \cite{Oli91,UO92,UO94}. By simplifying the event rules we 
obtain a set of semantically equivalent rules, but with a lower evaluation 
cost. The automatic generation and simplification of the event rules has been 
implemented in a SunOS environment by means of Quintus Prolog. In this paper, 
we will consider the simplified version of the event rules.

\begin{definition} Let D = (EDB, IDB) be a deductive database. The 
\textit{augmented database} associated to D is the tuple A(D) = (EDB, IDB*), where IDB* contains the 
rules in DR $ \cup $ IC $ \cup $ Cond and their associated simplified 
transition and event rules.
\end{definition}

\begin{example}
\label{exEventRules}
Given the following database D = (EDB,IDB):

\begin{tabbing}
XXXXX\=\kill
\>Sign(John) \\
\>Fail-ex(John) \\
\>Cont(x) $ \leftarrow $ Sign(x) $ \wedge \neg$Fail-ex(x)
\end{tabbing}
\noindent
the corresponding augmented database A(D) is the following:

\begin{tabbing}
XXXXX\=\kill
\>Sign(John) \\
\>Fail-ex(John) \\
\>Cont(x) $ \leftarrow $ Sign(x) $ \wedge \: \neg$Fail-ex(x) \\

\>$\iota $Cont(x) $ \leftarrow $ Sign(x) $ \wedge \: \neg \delta 
$Sign(x) $ \wedge \: \delta $Fail-ex(x) \\

\>$\iota $Cont(x) $ \leftarrow \iota $Sign(x) $ \wedge \: \neg$ Fail-ex(x) $ \wedge \: \neg \iota $Fail-ex(x) \\

\>$\iota $Cont(x) $ \leftarrow \iota $Sign(x) $ \wedge \: \delta $Fail-ex(x) \\

\>$\delta $Cont(x) $ \leftarrow $ Cont(x) $ \wedge \: \iota $Fail-ex(x) \\

\>$\delta $Cont(x) $ \leftarrow \delta $Cont(x) $ \wedge \: \neg$ Fail-ex(x)
\end{tabbing}
\end{example}

\subsection{Using other Rules Instead of the Event Rules}
\label{sectOtherRules}

We will use the event rules to provide the basis of our framework for 
specifying and handling schema validation and update processing problems. 
However, since our framework is only based on the definition of event given 
in (1) and (2) (see Section 3.1), we could use any set of rules that defines 
the difference between consecutive states of the database, instead of the 
event rules, provided that they preserve the event definition. In this 
section we show two different sets of rules that could had been used also. 

Assume that we have the following deductive rules:

\begin{tabbing}
XXXXX\=\kill
\>P(x) $ \leftarrow $ Q(x) $ \wedge $ R(x) \\
\>R(x) $ \leftarrow $ S(x)
\end{tabbing}

If we simply consider the definition of transition and event rules 
without applying any simplification, we would have:

\begin{tabbing}
XXXXX\=\kill
\>$\iota $P(x) $ \leftarrow P^{n}$(x) $ \wedge \: \neg$P(x) \\

\>$\iota $R(x) $ \leftarrow R^{n}$(x) $ \wedge \: \neg$R(x) \\

\>$\delta $R(x) $ \leftarrow $ R(x) $ \wedge \: \neg$R$^{n}$ (x) \\

\>P$^{n}$(x) $ \leftarrow $ Q(x) $ \wedge  \: \neg \delta $Q(x) $ \wedge 
$ R(x) $ \wedge  \: \neg \delta $R(x) \\

\>P$^{n}$(x) $ \leftarrow $ Q(x) $ \wedge  \: \neg \delta $Q(x) $ \wedge 
\: \iota $R(x) \\

\>P$^{n}$(x) $ \leftarrow  \: \iota $Q(x) $ \wedge $ R(x) $ \wedge $ 
$\neg \delta $R(x) \\

\>P$^{n}$(x) $ \leftarrow  \iota $Q(x) $ \wedge \: \iota $R(x) \\

\>R$^{n}$(x) $ \leftarrow $ S(x) $ \wedge  \: \neg \delta $S(x) \\

\>R$^{n}$(x) $ \leftarrow  \iota $S(x)
\end {tabbing}

On the other hand, by adapting the rules generated by K\"{u}chenhoff \cite{Kuc91} to our terminology, we would obtain: 

\begin{tabbing}
XXXXX\=\kill
\>$\iota $P(x) $ \leftarrow $ Q(x) $ \wedge  \: \neg \delta $Q(x) $ 
\wedge \: \iota $R(x) $ \wedge \: \neg$P(x) \\

\>$\iota $P(x) $ \leftarrow \iota $Q(x) $ \wedge $ R(x) $ 
\wedge  \: \neg \delta $R(x) $ \wedge \: \neg$P(x) \\

\>$\iota $P(x) $ \leftarrow \iota $Q(x) $ \wedge \: \iota 
$R(x) $ \wedge \: \neg$P(x) \\

\>$\iota $R(x) $ \leftarrow \: \iota $S(x) \\

\>$\delta $R(x) $ \leftarrow \: \delta $S(x)
\end{tabbing}

Note that K\"{u}chenhoff's rules provide some simplifications with regards to 
the non-simplified event rules. For instance, insertion event rules about R do 
not check that R was false in the old state of the database. A similar 
simplification is given for deletion event rules about R. However, no 
simplification is applied for the rules defining events on P.

Finally, the simplified insertion event rules for P according to \cite{UO92} are the following:

\begin{tabbing}
XXXXX\=\kill
\>$\iota $P(x) $ \leftarrow $ Q(x) $ \wedge  \: \neg \delta $Q(x) $ 
\wedge \: \iota $R(x) \\

\>$\iota $P(x) $ \leftarrow \iota $Q(x) $ \wedge $ R(x) $ 
\wedge  \: \neg \delta $R(x) \\

\>$\iota $P(x) $ \leftarrow \iota $Q(x) $ \wedge \: \iota $R(x) \\

\>$\iota $R(x) $ \leftarrow \iota $S(x) \\

\>$\delta $R(x) $ \leftarrow \delta $S(x)
\end{tabbing}

Note that, in addition to the simplifications already provided by K\"{u}chenhoff, these rules include also simplifications involving the event rules for P.

It is not difficult to see that the three sets of rules are semantically 
equivalent and define the same transitions between consecutive database 
states. The event definition is preserved in all cases. The only difference 
relies on the kind of optimizations that have been considered. In fact, we 
could also think about other sets of rules that incorporate additional 
optimizations (see for instance \cite{UO94} which incorporates the knowledge 
provided by the integrity constraints into such set of rules). The 
differences among possible sets of rules imply advantages or inconveniences 
as far as efficiency is concerned, but not regarding the ability of solving 
the problems we deal with in this paper. Thus, our framework does not depend 
on any particular set of rules.

\section{Reasoning on the Event Rules}

There is a big amount of problems related to database schema validation and 
to update processing. Unfortunately, the general approach considered by 
previous research in this area has been to deal with each problem in an 
isolated way. So, it is unusual to find a method able to handle several of 
the previous problems. This limitation can be overcome by considering a set 
of rules that explicitly define the difference between two consecutive 
database states and by performing deductive and abductive reasoning on these 
rules.

The role of deduction and abduction as reasoning paradigms is widely 
accepted. \textit{Deduction} is an analytic process based on the application of general rules 
to particular cases, with the inference of a result. \textit{Abduction} is another form of 
synthetic reasoning which infers the case from the rule and the result.

The event rules define the exact changes, either on base as on derived 
predicates, produced as a consequence of the application of a given 
transaction to a database state. Deductive and abductive reasoning can be 
performed on those rules. As we will see, performing deductive reasoning on 
the event rules defines changes on derived predicates induced by changes on 
base predicates. On the other hand, performing abductive reasoning on the 
event rules defines changes on base predicates needed to satisfy changes on 
derived predicates. In this way, reasoning 
deductively or abductively on the event rules provides the basis for solving 
database schema validation and update processing problems in a uniform way. 

In fact, as stated in Section \ref{sectOtherRules}, any set of rules that precisely 
defines the difference between consecutive database states could also be used. 
Due to our previous experience with the event rules, we have considered them 
in this paper.

\subsection{Reasoning Deductively on the Event Rules}

Deduction is concerned with inferring consequences from facts via deductive 
rules. For instance, given a deductive rule P(x) $ \leftarrow $ Q(x) and a 
fact Q(A), deduction infers P(A) as a consequence of Q(A). Thus, deductive 
reasoning is suitable among other things for finding correct answers to 
queries.

\begin{definition}
Let D = (EDB, IDB) be a deductive database and G a 
goal L$_{1} \wedge \ldots \wedge L_{n}$. A \textit{correct answer} to G over EDB is a 
substitution $\theta $ for variables of G such that G$\theta $ is a logical 
consequence of EDB $ \cup $ IDB, i.e. EDB $ \cup $ IDB $\models$ G$\theta $.
\end{definition}

Since event rules define the changes that occur in a transition from an old 
state of a database to a new one as a consequence of the application of a 
given transaction, by considering deduction in the context of the augmented 
database we can also define how to reason deductively on the event rules.

\begin{definition}
Let D = (EDB, IDB) be a deductive database, A(D) = 
(EDB, IDB$^{\ast })$ its corresponding augmented database, T a transaction 
consisting of a set of ground base event facts, $ u $ a derived event. The\textit{ deduced consequences on} $u$ due to 
the application of T is the set $\theta $ of correct answers to EDB $ \cup $ 
IDB$^{\ast } \:  \cup $ T $ \cup  \: u$. Note that each correct answer $\theta 
_{i } \in \: \theta $ defines a ground derived event $u\theta _{i}$ induced as a consequence of the application of T.
\end{definition}

Thus, reasoning deductively on the event rules defines changes on derived 
predicates induced by changes on base predicates, since $\theta $ defines 
all ground facts about $u$ induced by the application of T to the current 
state of the database.

As an example, given the database of Example \ref{exEventRules} and the transaction 
T={\{}$\delta $Fail-ex(John){\}}, it is not difficult to see that reasoning 
deductively on the event rules allows to deduce that T induces $\iota 
$Cont(John), i.e. $\theta $ ={\{}x=John{\}}.

\subsection{Reasoning Abductively on the Event Rules}

Abduction is aimed at looking for hypothesis that explain a given 
observation, according to known laws usually specified by means of deductive 
rules. Abduction (in the absence of integrity constraints) is traditionally 
defined as follows: Given a set of sentences $T$ (a theory presentation) and a 
sentence $G $ (observation), the abductive task consists of finding a set of 
sentences $\Delta $ (abductive explanation for $G)$ such that: 

\begin{tabbing}
XXXXX\=\kill
\>(1) $T \: \cup \: \Delta \models \: G$
\end{tabbing}

It is usually considered that $\Delta $ consists of atoms drawn from 
predicates explicitly indicated as abducible (those whose instances can be 
assumed when required). Therefore, an \textit{abductive framework} is a pair 
$\prec \: T, \textit{Ab} \succ$, where \textit{Ab} is the set of 
abducible predicates, i.e. $\Delta  \:  \subseteq $ \textit{Ab}\footnote{ Here and in 
the rest of the paper we will use \textit{Ab} to indicate both the set of abducible 
predicates and the set of all their variable-free instances.}.

By considering these ideas in the context of the augmented database, we can 
also define how to reason abductively on the event rules. In this case, 
abducible predicates correspond to base event facts since this is the only 
possible way to physically update a database.

\begin{definition}
Let us consider a deductive database D = (EDB, IDB) 
and its corresponding augmented database A(D) = (EDB, IDB$^{\ast })$. We can 
define an associated abductive framework $\prec$ EDB $ \cup $ IDB$^{\ast }$, 
\textit{Ab}$ \: \succ$, where \textit{Ab} corresponds to the set of base event predicates. Now, given a 
ground derived event $u$, we can define an \textit{abductive explanation for} $u$ in $\prec$ EDB $ \cup $ IDB$^{\ast }$, \textit{Ab}$ \: \succ$
to be any set T$_{i}$ consisting of ground facts about predicates in \textit{Ab} such that:

\begin{tabbing}
XXXXX\=\kill
\>- EDB $ \cup $ IDB$^{\ast } \:  \cup $ T$_{i} \: \models \: u$
\end{tabbing}
\end{definition}

An explanation T$_{i}$ is \textit{minimal} if no proper subset of T$_{i}$ is also an 
explanation, i.e. if it does not exist any explanation T$_{j}$ for $u$ such 
that T$_{j } \subset $ T$_{i}$. 

The previous condition states that the update request is a logical 
consequence of the database updated according to T$_{i}$. Thus, abductive 
reasoning on the event rules defines changes on base predicates needed to 
satisfy a given change on a derived predicate.

As an example, given the database of Example \ref{exEventRules} and the derived event 
$\iota $Cont(John), it is not difficult to see that T={\{}$\delta 
$Fail-ex(John){\}} is a minimal abductive explanation for $\iota 
$Cont(John). That is, the insertion of Cont(John) is satisfied by the 
deletion of Fail-ex(John).

In general, the result of applying abductive reasoning may not be unique. 
That is, several sets T$_{i}$ of base event facts that satisfy a derived 
event may exist. Each possible set T$_{i}$ constitutes a possible 
transaction that applied to the database will accomplish the desired change 
on the derived predicate. Minimal explanations are usually of particular 
interest, specially when we deal with database updating problems since they 
allow to minimize the difference between the old and the new states of the 
database.

If no solution T$_{i}$ is obtained then either the requested update cannot 
be satisfied only by changes on the EDB or the current database state 
already satisfies the intended effect of the request (e.g. an insertion of P 
is requested into a database that already satisfies P). Note that in the 
latter case we do not obtain a solution with the empty set since, when 
taking the event definition into account, an insertion of P can only be 
explained if the old state of the database does not satisfy P (resp. a 
deletion of P can only be explained if P is satisfied).

With this framework, the basic strategy of a proof procedure for computing 
T$_{i}$ is the following. Given, for instance, an update request for 
inserting P, the update procedure tries to solve abductively the goal $ 
\leftarrow \iota $P in 
$\prec$ EDB $ \cup $ IDB$^{\ast }$, \textit{Ab}$ \: \succ$
generating a 
set T$_{i}$ of abducibles such that T$_{i}$ satisfies the above condition. 
Before abducing a base event, we have to check that its definition is 
satisfied. That is, for abducing an event $\delta $P we require P to be true 
in the old state of the database while for $\iota $P we require P to be 
false. The set T$_{i}$ generated by abduction for an update request $u$ 
constitutes a transaction that, applied to the current state of the EDB, 
will satisfy $u$.

Given an event (base or derived) u$_{i}$ to be explained, abductive 
reasoning on the event rules can also be used to determine sets of base 
events that ensure that a certain derived event u$_{j}$ is not induced by 
the explanations of u$_{i}$. In this case, the abductive interpretation 
defines changes on base predicates needed to satisfy that a certain change 
on a derived predicate does not occur as a consequence of the application of 
the explanations of u$_{i}$.

\begin{definition}
Let D be a deductive database D = (EDB, IDB), its corresponding augmented 
database A(D) = (EDB, IDB$^{\ast })$ and its associated abductive framework 
$\prec$ EDB $ \cup $ IDB$^{\ast }$, \textit{Ab}$ \: \succ$. Now, given a 
positive event $u_{i}$ and a negative derived event $\neg u_{j}$ such that
$u_{i} \wedge \: \neg u_{j}$ is allowed, we 
can define the abductive explanation for $u_{i} \wedge \: \neg u_{j}$ in 
$\prec$ EDB $ \cup $ IDB$^{\ast }$, \textit{Ab}$ \: \succ$ to be any set 
$T_{i}$
consisting of ground facts about predicates in \textit{Ab} such that:

\begin{tabbing}
XXXXX\=\kill
\>- EDB $ \cup  \: IDB^{\ast } \:  \cup \: T_{i} \models u_{i}$ \\

\>- EDB $ \cup  \: IDB^{\ast } \:  \cup \: T_{i} \not \models u_{j}$
\end{tabbing}

The first condition states that u$_{i}$ is a logical consequence of the 
database updated according to T$_{i}$, while the second states that u$_{j}$ 
it is not. Note that if the explanations of u$_{i}$ alone do not induce 
u$_{j}$, then they are already valid abductive explanations. 
\end{definition}

As an example, given the database of Example \ref{exEventRules} and the positive event 
$\iota $Sign(Mary) and the negative event $\neg \iota $Cont(Mary), we have 
that T={\{}$\iota $Sign(Mary), $\iota $Fail-ex(Mary){\}} is a minimal 
abductive explanation for $\iota $Sign(Mary) $ \wedge \: \neg \iota 
$Cont(Mary). Note that $\iota $Sign(Mary) alone would induce $\iota 
$Cont(Mary). However, adding $\iota $Fail-ex(Mary) to T does not induce 
$\iota $Cont(Mary) any more since Cont(Mary) will be false in the new 
database state.

Two special cases are of particular interest. First, when the negative 
derived event is $\neg \iota $Ic it is guaranteed that the obtained 
explanations do not induce any insertion of an integrity constraint. Thus, 
the new database state will be consistent if the old database state was 
already consistent. Second, if the positive event is base (i.e. a 
transaction T), abductive reasoning on the event rules determines possible 
sets S$_{i}$ of ground base event facts that, appended to T, ensure that the 
application of any of the resulting transactions T$_{i}$ = S$_{i} \: 
\cup $ T does not induce u$_{j}$. Note that if T alone does not induce 
u$_{j}$, then T itself is a valid transaction.

This framework can be easily generalized to reason abductively on sets of 
positive and negative events.

\section{Deductive or Abductive Nature of Database Problems}
\label{sectFramework}

Deduction and abduction provide a uniform way to reason about the event 
rules and, in general, about any set of rules that explicitly define the 
exact difference between two consecutive database states. Moreover, either 
views (i.e. derived predicates) or integrity constraints or conditions to be 
monitored are uniformly defined by means of deductive rules and they are 
only distinguished by the different semantics endowed to the head of the 
rule. Thus, a view defines common knowledge shared by different users, an 
integrity constraint defines a situation that must never happen and a 
condition to be monitored defines an information whose changes must be 
reported to the user.

Therefore, given a derived predicate P(x) defined by the rule P(x) $ 
\leftarrow $ Q(x) $ \wedge \: \neg$R(x), P can be expressed as:

\begin{tabbing}
XXXXX\=\kill
\>View(x) $ \leftarrow $ Q(x) $ \wedge \: \neg$R(x) \\

\>Ic1(x) $ \leftarrow $ Q(x) $ \wedge \: \neg$R(x) \\

\>Cond(x) $ \leftarrow $ Q(x) $ \wedge \: \neg$R(x)
\end{tabbing}

\noindent
according to the concrete semantics that we would like to endow to P.

Now, reasoning deductively or abductively on the event rules corresponding 
to View, Ic1 and Cond we may classify as naturally deductive or naturally 
abductive the database schema validation and update processing problems.

This is summarized in Table \ref{taula}. Each row corresponds to the form of 
reasoning to be applied to the event rules of P and to the relevant events 
about P (i.e. $\iota $P, $\delta $P, T $ \wedge  \: \neg \iota $P or T $ 
\wedge  \: \neg \delta $P; being T a transaction) to reason about. Each 
column considers a different semantics to be endowed to P. Finally, each 
resulting cell defines a possible database schema validation or update 
processing problem that can be specified in terms of that form of reasoning 
and of the considered semantics.

\begin{table}
\centerline{\includegraphics[scale=0.8]{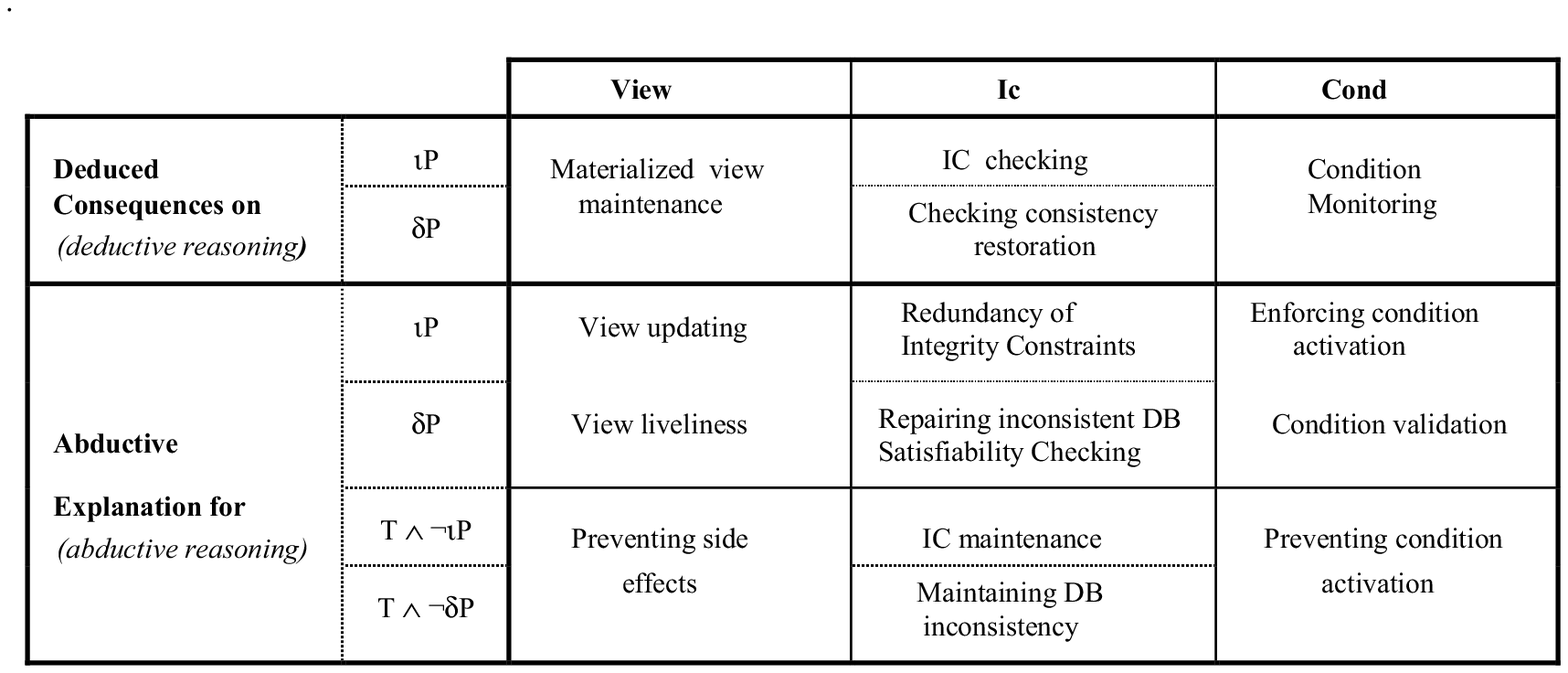}}
\caption{Classification of Database Problems}
\label{taula}
\end{table}

In the rest of this section, we briefly review database schema validation 
and update processing problems and we show how they can be handled by means 
of deductive or abductive reasoning, according to the classification 
provided in Table \ref{taula}.

\subsection{Schema Validation and Update Processing Problems}

The correct use of a database involves three different tasks: \textit{Schema Validation}, to guarantee 
that the database schema satisfies the user's intended needs and 
requirements; \textit{Query Processing}, to be able to give efficient and correct answers to the 
user' queries; and \textit{Update Processing}, to be able to correctly perform updates to the database 
contents. In general, several problems may arise when dealing with each of 
these tasks \cite{Ten00}. We will consider here only the problems encountered 
during schema validation and update processing, since query processing 
is beyond the scope of this paper.

\begin{example}
\label{exvalidation}
Consider the following flawed database schema to be validated:

\begin{tabbing}
XXXXX\=XXXXXXXXXXX\=\kill
\>(DR.1) \> Some-cand $ \leftarrow $ Cand(x) \\

\>(DR.2) \> Emp(x) $ \leftarrow $ Cand(x) $ \wedge $ Cont(x) \\

\>(DR.3) \> Cont(x) $ \leftarrow $ Sign(x) $ \wedge \: \neg$Fail-ex(x) \\

\>(DR.4) \> App(x) $ \leftarrow $ Cand(x) \\

\>(IC.1) \> Ic1(x) $ \leftarrow $ App(x) $ \wedge $ Sign(x) \\

\>(IC.2) \> Ic2(x) $ \leftarrow $ App(x) $ \wedge \: \neg$Has-account(x) \\

\>(IC.3) \> Ic3(x) $ \leftarrow  \: \neg $Some-cand \\

\>(IC.4) \> Ic4(x) $ \leftarrow $ Cand(x) $ \wedge \: \neg$App(x) \\

\>(IC.5) \> Ic5(x) $ \leftarrow $ App(x) \\

\>(IC.6, ..., IC.10) \> Ic $ \leftarrow $ Ici, for i=1...5 \\

\>(Cond.1) \> Cond1(x) $ \leftarrow $ Cand(x) $ \wedge \: \neg$Cont(x) \\

\>(Cond.2) \> Cond2(x) $ \leftarrow $ Emp(x) $ \wedge \: \neg$Cont(x)
\end{tabbing}

This schema defines four derived predicates (through deductive rules DR.1 to 
DR.4): Some-cand, Emp (employee), Cont (contracted person) and App 
(applicant). A person is an applicant if he is a candidate (Cand). A person 
is contracted if he signed an agreement (Sign) and he did not fail the exam. 
Employees are candidates that have a contract. Finally, Some-cand is true if 
the database contains, at least, one candidate.

The schema contains also five integrity constraints (defined by integrity 
rules IC.1 to IC.5). Ic1 states that it is not possible to have applicants 
that have signed an agreement. Ic2 states that it is not possible to be 
applicant and not to have an account. Ic3 states that there must be some 
candidate. Ic4 states that it is not possible to be candidate and not be 
applicant. Finally, Ic5 states that the database may not contain any 
applicant.

The schema contains also two conditions to be monitored. The first one is 
used to notify changes on the populations of applicants that do not have a 
contract and the second one changes on the populations of employees that do 
not have a contract.
\end{example}

\subsection{Schema Validation Problems}

In general, we cannot be completely sure that a certain database schema 
adequately describes the structure of the information that we want the 
database to contain. At first glance, we could perhaps detect that a certain 
deductive rule or integrity constraint is not precisely defined, as it might 
happen with IC.5 above, but it is very difficult to assess whether a certain 
schema does not present critical flaws. Detecting and removing flaws during 
schema design time will prevent these flaws from materializing as run-time 
errors or other inconveniences during operation time. \cite{DTU96} identified 
several desirable properties that a database schema should satisfy.

\subsubsection{Satisfiability Checking}

A database schema is \textit{satisfiable} if there exists an EDB for which no integrity 
constraint is violated \cite{BM86}, also mentioned in \cite{BDM88,IKH92}. Clearly, a 
non-satisfiable schema is not useful since it does not accept any 
extensional information.

As an example, the previous database schema is not satisfiable for any EDB. 
The empty EDB is not a proper EDB since it violates Ic3. Then, we need to 
consider an EDB with at least one candidate, let us say John. However, this 
is not enough since Ic4 would then be violated. So, John must also be an 
applicant but this is not possible since Ic5 impedes it. As a consequence of 
detecting that the previous schema is not satisfiable, we assume that the 
database designer decides to discard Ic3 and Ic5.

\textit{Satisfiability checking} can be naturally specified as performing abductive reasoning on the event 
rules associated to $\delta $Ic provided that Ic holds with an empty EDB. If 
there exists at least one abductive explanation for $\delta $Ic in 
$\prec$ EDB $ \cup $ IDB$^{\ast }$, \textit{Ab}$ \: \succ$,
with IDB = DR $ \cup $ IC, then the integrity 
constraints are satisfiable. Note that if Ic does not hold in the state 
corresponding to the empty EDB, all constraints are already satisfied in 
that state.

Note that, since satisfiability checking is to be determined at schema 
validation time, we are considering the empty EDB for checking this 
property. For the same reason, we will also use the empty EDB for checking 
other problems related to database schema validation.

\subsubsection{Absolute Redundancy of an Integrity Constraint}

Intuitively, an integrity constraint is absolutely redundant if integrity 
does not depend on it. That is, if it can never be violated. Obviously, an 
absolute redundant integrity constraint is not useful since it does not add 
any additional information to the information already provided by the rest 
of the schema.

For instance, integrity constraint Ic4 is absolutely redundant since the 
deductive rule DR.4 prevents Ic.4 to be violated for any EDB. Therefore, the 
database designer has to modify the schema to remove this absolute 
redundancy. In this case, we assume that he decides to discard DR.4.

Given an integrity constraint $Ic_{i}$, \textit{absolute} \textit{redundancy} can be naturally specified as performing abductive reasoning on the event 
rules associated to $\iota Ic_{i}$. If there exists at least one abductive 
explanation for $\iota Ic_{i}$ in 
$\prec$ EDB $ \cup $ IDB$^{\ast }$, \textit{Ab}$ \: \succ$,
with IDB = DR $ \cup $ 
IC, then $Ic_{i}$ is not absolutely redundant since it can be violated in some 
state of the database. In particular, in the database state that we obtain as 
a result of applying the obtained abductive explanation.

\subsubsection{View Liveliness}

A derived predicate P (i.e. a view) is \textit{lively} if there exists an EDB in which at 
least one fact about P is true. That is, predicates which are not lively 
correspond to views that are empty in each possible state of the database. 
Such predicates are clearly not useful and probably ill-specfied. This 
definition of ``liveliness'' essentially coincides with the definition of 
``satisfiable'' in \cite{LS95}.

For instance, predicate Emp as defined in Example \ref{exvalidation}
 is not lively. The 
reason is that a fact Emp(X) requires Cand(X) and Sign(X) to be true at the 
same time. However, since nobody can be a candidate without being an 
applicant (Ic.4) and nobody can be an applicant and to have signed an 
agreement (Ic.1), no person X can be an employee. We assume that the 
database designer decides to correct this flaw by redefining Ic.1 as 
Ic1'(x) $ \leftarrow $ App(x) $ \wedge $ Sign(x) $ \wedge \: \neg$ Has-account(x).

In our framework, provided that no fact about View holds in the empty EDB, 
\textit{view liveliness} can be specified as reasoning abductively on the event rules of $\iota 
$View(\textbf{x}). If there exists at least one abductive explanation for a 
certain event $\iota $View(\textbf{X}) in 
$\prec$ EDB $ \cup $ IDB$^{\ast }$, \textit{Ab}$ \: \succ$,
then View is 
lively since it is possible to reach a state where at least a fact 
View(\textbf{X}) is true. Otherwise, View is not lively. Note that if some 
fact about View holds in the empty EDB, then View is already lively in that 
state and there is no reason to ask about this property.

Let us review the Example \ref{exvalidation} with the flaws corrected up to 
now:

\begin{tabbing}
XXXXX\=XXXXXXXXXXXX\=\kill
\>(DR.1) \> Some-cand $ \leftarrow $ Cand(x) \\

\>(DR.2) \> Emp(x) $ \leftarrow $ Cand(x) $ \wedge $ Cont(x) \\

\>(DR.3) \> Cont(x) $ \leftarrow $ Sign(x) $ \wedge \: \neg$Fail-ex(x) \\

\>(IC.1') \> Ic1(x) $ \leftarrow $ App(x) $ \wedge $ Sign(x)$ \wedge 
\: \neg$Has-account(x) \\

\>(IC.2) \> Ic2(x) $ \leftarrow $ App(x) $ \wedge \: \neg$ Has-account(x) \\

\>(IC.4) \> Ic4(x) $ \leftarrow $ Cand(x) $ \wedge \: \neg$App (x) \\

\>(IC.6,IC.7, IC.9) \> Ic $ \leftarrow $ Ici, for i=1,2, 4. \\

\>(Cond.1) \> Cond1(x) $ \leftarrow $ Cand(x) $ \wedge \: \neg$Cont (x) \\

\>(Cond.2) \> Cond2(x) $ \leftarrow $ Emp(x) $ \wedge \: \neg$Cont (x)
\end{tabbing}

\subsubsection{Relative Redundancy of Integrity Constraints}

Relative redundancy is similar to absolute redundancy but, in this case, an 
integrity constraint (or a set of constraints) is relatively redundant if it 
is always satisfied in all states that satisfy the rest of the constraints. 
Again, such a redundancy should be detected and redundant constraints should 
not be considered during update processing.

In our example, we can see that Ic1' is relatively redundant since it is 
entailed by Ic2. Therefore, we assume that the database designer decides to 
discard Ic.1' since the resulting database will admit the same consistent 
states.

Given an integrity constraint $Ic_{i}$, \textit{rlative} \textit{redundancy} can be naturally specified as performing abductive reasoning on the event 
rules associated to $\iota Ic_{i}$. $Ic_{i}$ is not relatively redundant if there 
exists at least one abductive explanation for $ \leftarrow \iota 
Ic_{i} \wedge  \: \neg \iota $Ic in 
$\prec$ EDB $ \cup $ IDB$^{\ast }$, \textit{Ab}$ \: \succ$,
with IDB = DR $ \cup $ IC $-$ {\{} Ic $ \leftarrow $ $Ic_{i}${\}}.

\subsubsection{Condition Validation}

Condition validation refers to the problem of determining whether it is 
possible to change the contents of a certain condition Cond(x). That is, to 
determine whether exists at least one transaction that, if applied to the 
database, could activate a certain condition $\iota $Cond(\textbf{X}) or 
$\delta $Cond(\textbf{X}). Clearly, a condition that can never be activated 
is probably ill-specified since the active behaviour of the database does 
not depend on it. This can be useful, for instance, to provide the database 
designer with a tool for validating certain aspects of the condition 
definition and, hence, of the active behaviour of the database. This problem 
is very important in the context of active databases since this technology 
is mainly based on the extensive use of conditions to be monitored, which 
are the core of Condition-Action (CA) and Event-Condition-Action (ECA) rules 
\cite{WC96}.

For instance, condition Cond2 as defined in Example \ref{exvalidation}
 is not valid since 
no insertion and no deletion can be induced on it. The reason is that, by 
the deductive rule DR.2, employees must have a contract and, then, it is not 
possible to have employees without a contract. So, we assume that the 
database designer decides to discard condition Cond2.

In a similar way that view liveliness, changes induced in a given condition, 
Cond(x), can be specified as reasoning abductively on the event rules of 
$\iota $Cond(\textbf{x}) and $\delta $Cond(\textbf{x}). If there exists at 
least one abductive explanation for a certain event $\iota $Cond(\textbf{X}) 
or $\delta $Cond(\textbf{X}) in 
$\prec$ EDB $ \cup $ IDB$^{\ast }$, \textit{Ab}$ \: \succ$,
then the condition can be activated.

\subsection{Update Processing Problems}
\label{subsectUpdProb}

Once the database schema is validated, we are ready to perform updates to 
the database contents. Several problems will arise when processing the 
requested updates \cite{TU95}. To illustrate these problems, we consider the 
schema we have previously validated and we will assume that the database 
contains several base facts. Note that, now, the schema is satisfiable, all 
predicates are lively and no integrity constraint is either absolutely nor 
relatively redundant.

\begin{example}
\label{exupdating}
The following database will be considered to deal 
with update problems related to views and integrity constraints:

\begin{tabbing}
XXXXX\=XXXXXXXXXX\=\kill
\>(F.1) \>Sign(John) \\

\>(F.2) \>Fail-ex(John) \\

\>(DR.1) \>Some-cand $ \leftarrow $ Cand(x) \\

\>(DR.2) \>Emp(x) $ \leftarrow $ Cand(x) $ \wedge $ Cont(x) \\

\>(DR.3) \>Cont(x) $ \leftarrow $ Sign(x) $ \wedge \: \neg$Fail-ex(x) \\

\>(IC.2) \>Ic2(x) $ \leftarrow $ App(x) $ \wedge \: \neg$Has-account(x) \\

\>(IC.4) \>Ic4(x) $ \leftarrow $ Cand(x) $ \wedge \: \neg$App (x) \\

\>(IC.7, IC.9) \>Ic $ \leftarrow $ Ici, for i=2, 4. \\

\>(Cond.1) \>Cond1(x) $ \leftarrow $ Cand(x) $ \wedge \: \neg$Cont (x)
\end{tabbing}
\end{example}

\subsubsection{Integrity Constraint Checking}

There exists a large cumulative effort in the field of \textit{integrity constraint checking} \cite{SK88,Kuc91,Oli91,GCM94,LL96,SJ96}. Given a consistent database and a transaction 
(i.e. a set of insertions and deletions of base facts), integrity constraint 
checking is devoted to incrementally, i.e. efficiently, determine whether 
the application of this transaction to the current database violates some 
integrity constraint. In this case, the transaction is rejected since, 
otherwise, its application would lead to an inconsistent database state.

Given a transaction T, \textit{integrity constraint checking} can be naturally specified in our framework as 
performing deductive reasoning on the event rules associated to $\iota $Ic, 
provided that Ic does not hold. The deduced consequences on $\iota $Ic are 
either the identity substitution or no correct answer of EDB $ \cup $ 
IDB$^{\ast } \:  \cup $ T $ \cup  \: \neg \iota $Ic exists. In the first case,
 T induces an insertion of Ic and, 
therefore, it must be rejected because it violates some integrity 
constraint. Otherwise, T does not violate any integrity constraint and it 
can be successfully applied. As it happens with materialized view 
maintenance, efficiency of the process is ensured since reasoning about the 
transaction and the event rules allows to compute only the updates induced 
by this transaction.

As an example, assume that the database of Example \ref{exupdating} contains 
also the 
facts App(Peter) and Has-account(Peter). Reasoning deductively on the event 
rules of Ic2 we could determine that the transaction T={\{}$\delta 
$Has-account(Peter){\}} induces the insertion of Ic2(Peter) and, thus, of Ic 
and would lead the database to an inconsistent state.

\subsubsection{Integrity Constraint Maintenance}

The main drawback of integrity constraint checking is that the user may not 
know which changes to the transaction are needed to guarantee that its 
application does not violate any integrity constraint. \textit{Integrity constraint maintenance} is aimed at 
overcoming this drawback: given a consistent database state and a 
transaction T that violates some integrity constraint, the problem is to 
find \textit{repairs}, that is, an additional set of insertions and/or deletions of base 
facts to be appended to T such that the resulting transaction T' satisfies 
all integrity constraints. In general, there may be several repairs and the 
user must select one of them. Eventually, if no such repair exists then the 
original transaction must be rejected. Several methods have been proposed to 
deal with this problem \cite{ML91,CFPT94,Wut93,ST94,TO95}.

Given a consistent database state and a transaction T, \textit{integrity constraint maintenance} can be specified in 
our framework as performing abductive reasoning on the goal $ \leftarrow 
T \:  \wedge  \: \neg \iota $Ic. Thus, possible abductive 
explanations for T $ \wedge  \: \neg \iota $Ic in 
$\prec$ EDB $ \cup $ IDB$^{\ast }$, \textit{Ab}$ \: \succ$,
with IDB = DR $ \cup $ IC, correspond to the possible transactions 
T', T $ \subseteq $ T', that maintain database consistency.

As an example, consider again the database of Example 
\ref{exupdating} and assume that 
the transaction T={\{}$\iota $App(Claire){\}} wants to be applied to the 
database. Reasoning abductively on $\iota $App(Claire) $ \wedge  \: \neg 
\iota $Ic we obtain the transaction T'={\{}$\iota $App(Claire), $\iota 
$Has-account(Claire){\}} which satisfies the original transaction and 
maintains the database consistent.

\subsubsection{View Updating}

\textit{View updating} is concerned with determining how a request to update a view, i.e. to 
update the contents of a derived predicate, must be appropriately 
translated into updates of the underlying base facts. In general, several 
translations may exist and the user must select one of them. This problem 
has attracted much research during the last years in deductive databases 
\cite{KM90,GL91,TO95,CST95,Dec96,LT97} and it has been already identified as 
an abductive problem \cite{CST95,Dec96,DD98,IS99}

\textit{View updating} can be naturally specified as performing abductive reasoning on the event 
rules of $\iota $View(\textbf{X}) or $\delta $View(\textbf{X}), where 
View(\textbf{X}) is the derived fact to be inserted or deleted, 
respectively. The abductive explanation for $\iota $View(\textbf{X}) defines 
possible sets of base fact updates (i.e. transactions) that satisfy the 
insertion of View(\textbf{X}), while the abductive explanation for $\delta 
$View(\textbf{X}) defines possible sets of base fact updates that satisfy 
the deletion of View(\textbf{X}).

For instance, in Example \ref{exupdating} reasoning abductively on the event 
rules of 
Cont we can determine that the view update request $\iota $Cont(John) is 
satisfied by the transaction T={\{}$\delta $Fail-ex(John){\}}.

In principle, it may happen that some translations corresponding to a given 
view update request do not satisfy the integrity constraints. For this 
reason, view updating is usually combined with problems related to integrity 
constraints. Possible ways of performing this combination 
will be explained in Section \ref{subsectCombining}.

\subsubsection{Materialized View Maintenance}

A view can be materialized by explicitly storing its contents in the 
extensional database. This can be useful, for instance, to improve 
efficiency of query processing. Given a transaction, \textit{materialized view maintenance} consists of 
incrementally, i.e. efficiently, determining which changes are needed to 
update accordingly the materialized views (see \cite{GM95,Ros98} for a 
state-of-the-art reports).

Given a transaction T and a materialized view View(\textbf{x}), 
\textit{materialized view maintenance} can be naturally specified as performing deductive reasoning on the event 
rules associated to $\iota $View(\textbf{x}) and $\delta 
$View(\textbf{x}). That is, deduced consequences for $\iota 
$View(\textbf{x}) and for $\delta $View(\textbf{x}) correspond, 
respectively, to the insertions and to the deletions to be performed on 
View(\textbf{x}). Efficiency of the process is ensured since reasoning about 
the transaction and the event rules allows to incrementally compute only the 
updates induced by this transaction.

For instance, if we assume that predicate Cont(x) in Example 
\ref{exupdating} is 
materialized, reasoning deductively on the event rules of Cont we can 
determine that the transaction T={\{}$\delta $Fail-ex(John){\}} induces the 
insertion of Cont(John) in the materialized view.

\subsubsection{Preventing Side Effects}

Due to the deductive rules, undesired updates may be induced on some derived 
predicates when applying a transaction. We say that a side effect occurs 
when this happens. The problem of \textit{preventing side effects} \cite{TO95} is concerned with determining a 
set of base fact updates which, appended to a given transaction, ensure that 
the application of the resulting transaction to the current state of the 
database will not induce the undesired side effects. In general, several 
solutions may exist and the user must select one of them.

Ensuring that a transaction T will not induce an insertion or a deletion of 
a derived fact View(\textbf{X}) can naturally be specified as reasoning 
abductively on {\{}T $ \wedge  \: \neg \iota $View(\textbf{X}){\}} or on 
{\{}T $ \wedge  \: \neg \delta $View(\textbf{X}){\}}, respectively. The 
former defines sets T' of base fact updates, which are supersets of T, 
needed to guarantee that the insertion of View(\textbf{X}) is not induced by 
T, while the latter defines sets T' of base fact updates, again supersets of 
T, needed to satisfy that the deletion of View(\textbf{X}) is not induced.

For instance, reasoning abductively on $\iota $Sign(Mary) $ \wedge  \: \neg 
\iota $Cont(Mary) we can prevent that the transaction T={\{}$\iota 
$Sign(Mary){\}} will not induce the insertion of Cont(Mary). This is done by 
considering T'={\{}$\iota $Sign(Mary), $\iota $Fail-ex(Mary){\}} instead of 
T, which is also given by this abductive interpretation.

\subsubsection{Condition Monitoring}

\textit{Condition monitoring} refers to the problem of incrementally monitoring the changes on a 
condition induced by a transaction that consists of a set of base fact 
updates \cite{RCB89,HCK90,QW91,BCP98}. 

As an example, applying the transaction T={\{}$\iota $Cand(Peter){\}} to the 
database of Example \ref{exupdating} would induce $\iota $Cond1(Peter).
That is, due to the application of T, Peter would be a candidate without a 
contract.

In our framework, changes induced in a condition Cond(x), are specified as 
performing deductive reasoning on the events rules associated to $\iota 
$Cond(\textbf{x}) and $\delta $Cond(\textbf{x}). The former, $\iota 
$Cond(\textbf{x}), defines the changes meaning that x satisfy the condition 
after the application of the transaction, but not before. The latter, 
$\delta $Cond(\textbf{x}), defines the changes meaning that x satisfy the 
condition before the application of the transaction, but not after.

\subsubsection{Enforcing Condition Activation}

\textit{Enforcing condition activation} refers to the problem of obtaining the possible transactions that, if 
applied to the current state of the database, would induce an activation of 
a given condition. For instance, the transaction T$_{1}$={\{}$\iota 
$Cand(Peter){\}} would induce the condition $\iota $Cond1(Peter).

In our framework, enforcing condition activation is specified reasoning 
abductively on $\iota $Cond(\textbf{X}) or $\delta $Cond(\textbf{X}), where 
both correspond to the conditions to be enforced. The former defines 
possible transactions that will induce \textbf{X} to satisfy the condition 
after their application, but not before. The latter, defines possible 
transactions that will induce \textbf{X} not to satisfy the condition after 
their application.

\subsubsection{Preventing Condition Activation}

This problem is close to the problem of preventing side effects but 
considering conditions to be monitored instead of views. Given a transaction 
T, the problem of \textit{preventing condition activation} is to find an additional set of insertions and/or 
deletions of base facts to be appended to T such that the resulting 
transaction T' guarantees that no changes in the condition would occur as a 
consequence of the application of T'$.$ In general, several resulting 
transactions may exist and the user should select one of them.

\subsection{Updates to an Inconsistent Database}

Sometimes it could be useful to allow for intermediate inconsistent database 
states, i.e. states where some integrity constraint is violated. This may 
happen, for instance, to reduce the number of times that integrity 
constraint enforcement is performed. In this case, three new problems 
related to update processing arise.

\subsubsection{Checking Consistency Restoration}

Given an inconsistent database state and a transaction that consists of a 
set of base fact updates, the problem of \textit{checking consistency restoration} is to incrementally check whether 
these updates restore the database to a consistent state.

\textit{Checking consistency restoration} can be specified as performing deductive reasoning on $\delta $Ic, provided 
that Ic holds. In this case, deduced consequences on $\delta $Ic are also 
either the identity substitution or no correct answer exists. If the 
identity substitution is obtained, then the transaction induces a deletion 
of Ic and, therefore, restores the database to a consistent state.

\subsubsection{Repairing an Inconsistent Database}

Given an inconsistent database state, the problem of \textit{repairing an inconsistent database} is to obtain a set of 
updates of base facts, i.e. a transaction, that restore the database to a 
consistent state. In general, several solutions may exist and the database 
administrator should select one of them.

The problem of \textit{repairing an inconsistent database} can be specified as performing abductive reasoning on the 
event rules associated to $\delta $Ic, provided that Ic holds. Given an EDB 
that violates some integrity constraint, abductive explanations for $\delta 
$Ic in 
$\prec$ EDB $ \cup $ IDB$^{\ast }$, \textit{Ab}$ \: \succ$,
with IDB = DR $ \cup $ IC, 
correspond to the possible transactions that would induce a deletion of Ic 
and that, therefore, would restore database consistency.

\subsubsection{Maintaining Database Inconsistency}

Given an inconsistent database state and a transaction T, the problem of 
\textit{maintaining database inconsistency} is to obtain an additional set of base fact updates to be appended to the 
original transaction to guarantee that the resulting database state remains 
inconsistent.

\textit{Maintaining database inconsistency} can be specified as performing abductive reasoning on the goal $ \leftarrow 
T \:  \wedge  \: \neg \delta $Ic, provided that Ic holds, with an 
abductive framework 
$\prec$ EDB $ \cup $ IDB$^{\ast }$, \textit{Ab}$ \: \succ$,
with IDB = DR $ \cup $ 
IC. Although we do not see for the moment any practical application of this 
problem, it can be naturally classified and specified in the framework we 
propose in this paper.

\subsection{Combining Different Problems}
\label{subsectCombining}

In previous sections, we have assumed that deductive or abductive reasoning 
is performed on the event rules associated to a single derived event 
predicate. However, this framework can be easily extended to consider 
several derived events instead of only one. Deductive or abductive reasoning 
on a set of derived events is performed by considering  the conjunction of 
all derived events in the set as the goal to be reasoned about. For 
instance a view update request consists, in general, of a set of insertions 
and/or deletions to be performed on derived predicates, e.g. $u=\iota $P(a) 
$ \wedge  \: \iota $Q(b)$ \wedge  \: \delta $S(c) stands for the 
request of inserting P(a) and Q(b) and deleting S(c), being P, Q and S 
derived predicates. In this case, translations that satisfy $u$ correspond to 
the abductive explanations for $\iota $P(a) $ \wedge  \: \iota 
$Q(b) $ \wedge  \: \delta $S(c) in 
$\prec$ EDB $ \cup $ IDB$^{\ast }$, \textit{Ab}$ \: \succ$.

Moreover, we would like to notice that deductive problems can be naturally 
combined. All of them share a common starting-point (a transaction that 
consists of a set of base fact updates) and aim at the same goal (to define 
the changes on derived predicates induced by this transaction). The same 
reasons allow the combination of abductive problems. Therefore, we 
can specify more complex database updating problems of deductive or of 
abductive by considering possible combinations of the problems specified in 
section \ref{sectFramework}.

For instance, given a transaction T, a materialized view View, a condition 
to be monitored Cond and the integrity constraint predicate Ic, we could 
combine materialized view maintenance, integrity constraints checking and 
condition monitoring by reasoning deductively on $\iota 
$View(\textbf{x}) $ \wedge  \: \delta $Cond(\textbf{y}) $ \wedge  \: \neg 
\iota $Ic. Deduced consequences on $\iota $View(\textbf{x}) $ \: \wedge $ 
$\delta $Cond(\textbf{x}) $ \wedge  \: \neg \iota $Ic correspond to the 
values \textbf{x} and \textbf{y} that cause an insertion of View, satisfy 
the condition $\delta $Cond as a consequence of the application of T, and 
such that T does not violate any integrity constraint.

In a similar way, we could also combine view updating with integrity 
constraints maintenance by reasoning abductively on $\iota $View(a) $
 \wedge  \: \neg \iota $Ic. In this case, abductive explanations for 
$\iota $View(a)$ \wedge  \: \neg \iota $Ic correspond to the 
translations that satisfy both the insertion of View(a) and that do not 
violate any integrity constraint.

Furthermore, we could also combine deductive and abductive problems. Note 
that the result of performing abductive reasoning is exactly the 
starting-point for performing deductive reasoning, that is, a transaction 
that consists of a set of base fact updates. Therefore, we could first deal 
with abductive problems and, immediately after, use the obtained result for 
dealing with the deductive ones.

For instance, we could be interested on distinguishing between integrity 
constraints to be maintained and integrity constraints to be checked, and on 
combining view updating with the treatment of both kinds of constraints. In 
this case, we should first reason abductively on the view update request and 
the set of integrity constraints to be maintained and, later on, to consider 
the resulting transactions and reason deductively on the set of integrity 
constraints to be checked to reject those resulting transactions that 
violate some constraint in this set.

Finally, we would also like to notice that in our approach for the 
specification 
of database updating problems does not change when considering other kinds of 
updates like insertions or deletions of deductive rules. In this case, we 
should first determine the changes on the transition and event rules caused 
by the update and apply then our approach.

\section{Using Existing Procedures to Reason on the Event 
Rules}

Our framework to classify database schema validation and updating processing 
problems is based on the existence of a set of rules, like the event rules, 
that define the exact difference between consecutive database states. By 
performing deductive and abductive reasoning on these rules, we can deal 
with all these problems in a uniform way. Therefore, our framework does not 
rely on the concrete method we use to perform either deductive or abductive 
reasoning. However, candidate methods to be used must satisfy several 
conditions.

Given a method able to perform deductive reasoning in a certain class of 
deductive databases, it should satisfy two requirements to tackle the 
deductive problems described in the previous section:

\begin{enumerate}
\item[a)]
The class of deductive databases considered by the method must allow 
expressing at least the goals required to define these problems.

\item[b)]
The method must obtain all correct answers that satisfy the request.
\end{enumerate}

Similarly, given a method able to perform abductive reasoning in a certain 
class of deductive databases, it should satisfy two requirements to tackle 
the abductive problems described in the previous section:

\begin{enumerate}
\item[a)]
The class of deductive databases considered by the method must allow 
expressing at least the goals required to define these problems.

\item[b)]
For schema validation problems, if there exists some explanation for a given 
request the method obtains such explanation (but not necessarily several or 
even all of them). For updating problems, the method must be complete, i.e. 
it must obtain all possible explanations that satisfy the request.
\end{enumerate}

In this section we show that there exists already some procedures able to 
compute each different form of reasoning on the event rules and we 
illustrate them by means of some examples. We show, in this way, the 
applicability of our approach. We would like to mention, however, that our 
aim is not that of comparing existing procedures but just to show they are 
able to handle several problems.

\subsection{Using SLDNF to Reason Deductively on the Event Rules}

Standard SLDNF resolution is a possible way for reasoning deductively on the 
event rules. Given an augmented database A(D) = (EDB, IDB$^{\ast })$, a 
transaction T and a derived event $u$, the deduced consequences on $u$ due to 
T correspond to the successful SLDNF derivations of the goal $ \leftarrow u $ 
that result in a computed answer $\theta _{i}$ when considering the 
input set EDB $ \cup $ IDB$^{\ast } \:  \cup $ T.

Nevertheless, other proof procedures could be used instead of SLDNF 
resolution like, for instance, bottom-up computation of the event rules. To 
motivate our discussion and without loss of generality, we will assume that 
SLDNF resolution is used to reason deductively on the event rules. The 
following example illustrates how to perform deductive reasoning on the 
event rules.

\begin{example}
\label{exDedReasoning}
Consider again Example \ref{exEventRules} and assume a transaction T 
which consists of the deletion of the base fact Fail-ex(John), in our 
notation T={\{}$\delta $Fail-ex(John){\}}. The corresponding deductive 
framework is:

\begin{tabbing}
XXXXX\=XXXXXXXXXX\=\kill
\>EDB $ \cup $ IDB$^{\ast }$: \>Sign(John) \\

\>\>Fail-ex(John) \\

\>\>Cont(x) $ \leftarrow $ Sign(x) $ \wedge \: \neg$Fail-ex(x) \\

\>\>$\iota $Cont(x) $ \leftarrow $ Sign(x) $ \wedge  \: \neg \delta 
$Sign(x) $ \wedge \: \delta $Fail-ex(x) \\

\>\>$\iota $Cont(x) $ \leftarrow \iota $Sign(x) $ \wedge  \: \neg $Fail-ex(x) 
$ \wedge  \: \neg \iota $Fail-ex(x) \\

\>\>$\iota $Cont(x) $ \leftarrow \iota $Sign(x) $ \wedge \: 
\delta $Fail-ex(x) \\

\>\>$\delta $Cont(x) $ \leftarrow \delta $Sign(x) $ \wedge \: \neg
$Fail-ex(x) \\

\>\>$\delta $ Cont(x) $ \leftarrow $Cont(x) $ \wedge \: \iota $Fail-ex(x) \\

\>T:\>$\delta $Fail-ex(John)
\end{tabbing}

\begin{figure}[hbt]
\centerline{\includegraphics{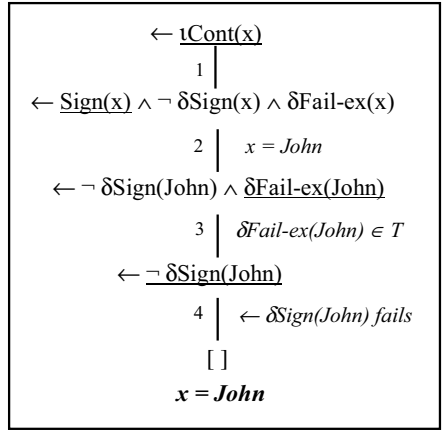}}
\caption{Computing answers for $ \leftarrow \iota $Cont(x)}
\label{figSLDNF}
\end{figure}

The SLDNF refutation of Figure \ref{figSLDNF} shows that T={\{}$\delta $Fail-ex(John){\}} 
induces the insertion of Cont(John), i.e. $\theta $ ={\{}x=John{\}}. Note that the SLDNF tree rooted at $ \leftarrow  \: \iota $Cont(x) succeeds 
with a computed answer x=John. That means that the deletion of Fail-ex(John) 
induces an insertion of Cont(John). Selecting other rules at step 1 of this 
tree does not result on any other successful branch and, thus, no 
other insertion of Cont is induced due to T.
\end{example}

\subsection{Using Abductive Procedures to Reason Abductively on the Event Rules}
\label{sectAbdProd}

We show now how the Events Method \cite{TO95}, Inoue and Sakama's method 
\cite{IS98,IS99} and SLDNFA \cite{DD98} may be used to perform abductive 
reasoning on the event rules and, hence, to deal with schema validation and 
updating processing problems. Other existing procedures could have been used 
as well like, for instance, \cite{KM90} which was the first attempt to use 
abduction in a database context. However, we have just considered some of the 
most recent proposals since they can be understood in some sense as an 
evolution of the initial ones.

A detailed discussion on the specific features and 
limitations of other (abductive) methods to perform view updating and 
integrity maintenance can be found in \cite{MT99}.

\subsubsection{The Events Method} 

The Events Method \cite{TO95} takes the event rules explicitly into account to obtain 
all possible minimal sets T$_{i}$ on the EDB that satisfy a given update 
request on the IDB. It extends the SLDNF proof procedure to obtain 
all possible transactions T$_{i}$ and it has been proved to be sound and 
complete for stratified databases \cite{TO95}. Soundness of the method guarantees 
that the obtained transactions satisfy the update request, while 
completeness ensures that it obtains all minimal transactions.

In this method, an update request $u$ is a conjunction of positive and 
negative events (base and derived). Positive events correspond to updates 
that must be effectively performed during the transition from the old state 
of the database to the new state, while negative events correspond to 
updates that may not happen during this transition.

Let D be a deductive database, A(D) its augmented database, $u$ an update 
request and T$_{i}$ a set of base events. In the Events Method, T$_{i}$ 
satisfies the request $u$ if, using SLDNF resolution, the goal $ \leftarrow  
u$ succeeds from input set A(D) $ \cup \: T_{i}$. Each set T$_{i}$ is 
obtained by having some failed SLDNF derivation of A(D) $ \cup  \:  
\leftarrow u$ succeed. The possible ways in which a failed derivation may 
succeed correspond to the different sets T$_{i}$ that satisfy the request. 
If no T$_{i}$ is obtained, then it is not possible to satisfy the requested 
update by changing only the EDB.

Although the event rules define the exact difference between consecutive 
database states, making a failed SLDNF derivation succeed does not always 
guarantee the generation of minimal solutions only. Therefore, the Events 
Method includes also a final step to discard obtained non-minimal solutions. 
We must note that the Events Method may not terminate in the presence of 
recursive rules because it may enter into an infinite loop. Moreover, as far 
as efficiency is concerned, it provides certain limitations on the treatment 
of rules with existential variables.

Let $u$ be an update request. In the Events Method, a transaction T satisfies 
$u$ if there is a constructive derivation from ($ \leftarrow u$ $\emptyset 
\: \emptyset)$ to ([ ] T C). The transaction T contains the base event facts 
to be applied, while the condition set C contains base events (subgoals in the 
general case) that would invalidate the update request if applied. A 
constructive derivation is defined as follows (for convenience, let G/L 
stand for the goal obtained from a goal G by dropping a selected occurrence 
of literal L in G):

\begin{definition} A \textit{constructive derivation} from (G$_{1}$ T$_{1}$ C$_{1})$ to (G$_{n}$ 
T$_{n}$ C$_{n})$ via a safe computation rule R \cite{Llo87} is a sequence: 

\begin{center}
(G$_{1}$ T$_{1}$ C$_{1})$, (G$_{2}$ T$_{2}$ C$_{2})$,\ldots , (G$_{n}$ 
T$_{n}$ C$_{n})$
\end{center}

\noindent
such that for each i $ \ge $ 1, G$_{i}$ has the form $ \leftarrow $L$_{1 } 
\wedge $\ldots $ \wedge \: $L$_{k}$, R(G$_{i})$ = L$_{j}$ and (G$_{i + 
1}$ T$_{i + 1}$ C$_{i + 1})$ is obtained according to one of the following 
rules:

\begin{enumerate}

\item[A1)]
If L$_{j}$ is a positive literal and it is not a base event,$_{ 
}$ then$_{ }$ G$_{i + 1}$= S, where S is the resolvent of some clause in A(D) 
with G$_{i}$ on the selected literal L$_{j}$, T$_{i + 1}$= T$_{i}$ and C$_{i 
+ 1}$= C$_{i}$.

\item[A2)]
If L$_{j}$ is a positive base event '$\iota $P' (resp. '$\delta $P'), 
there is a substitution $\sigma $ such that P$\sigma $ does not hold (resp. 
P$\sigma $ holds) in the current database and there is a consistency 
derivation from (C$_{i}$ T$_{i} \cup ${\{}L$_{j}\sigma ${\}} C$_{i})$ to 
({\{}{\}} T' C') then G$_{i + 1}$=G$_{i}\sigma 
$/L$_{j}\sigma $, T$_{i + 1}$=T' and C$_{i + 1}$= C'.

Note that if C$_{i }$= {\O} or L$_{j}\sigma  \in $ T$_{i}$ then G$_{i + 1 
}$= G$_{i}\sigma \backslash $L$_{j}\sigma $, T$_{i + 1 }$=T$_{i} 
\cup ${\{}L$_{j}\sigma ${\}} and$_{ }$C$_{i + 1}$= C$_{i}$.

\item[A3)]
If L$_{j}$ is negative and there is a consistency derivation from ({\{}$ 
\leftarrow \neg $L$_{j}${\}} T$_{i}$ C$_{i})$ to ({\{}{\}} T' C'), then 
G$_{i + 1}$=G$_{i}$/L$_{j}$, T$_{i + 1}$=T' and C$_{i + 1}$= C'.

\end{enumerate}
\end{definition}

Rule A1) is an SLDNF resolution step where A(D) acts as input set. In rule 
A2), the selected base event is included in the transaction set T$_{i}$ to 
get a successful derivation for the current branch, provided that the event 
does not violate any of the conditions in C$_{i}$. In rule A3), we get the 
next goal if we can ensure consistency for the selected literal.

\begin{definition} A \textit{consistency derivation} from (F$_{1}$ T$_{1}$ C$_{1})$ to (F$_{n}$ 
T$_{n}$ C$_{n})$ via a safe computation rule R is a sequence: 

\begin{center}
(F$_{1}$ T$_{1}$ C$_{1})$, (F$_{2}$ T$_{2}$ C$_{2})$,\ldots , (F$_{n}$ 
T$_{n}$ C$_{n})$
\end{center}

\noindent
such that for each i $ \ge $ 1, F$_{i}$ has the form H$_{i} \:  \cup $ 
F'$_{i}$, where H$_{i }= \leftarrow $L$_{1 } \wedge $\ldots $ \wedge 
\: $L$_{k}$ and, \textit{for some} j=1\ldots k, (F$_{i + 1}$ T$_{i + 1}$ C$_{i + 1})$ is 
obtained according to one of the following rules:

\begin{enumerate}

\item[B1)]
If L$_{j}$ is a positive literal and it is not a base event, then$_{ 
}$F$_{i + 1}$= S'$ \cup $F'$_{i}$, T$_{i + 1}$= T$_{i}$ and C$_{i + 1}$= 
C$_{i}$; where S' is the set of all resolvents of clauses in A(D) with 
H$_{i}$ on the selected literal L$_{j}$ and []$ \notin $S'. Note that, if no 
input clause in A(D) can be unified with L$_{j}$, then S' = $\emptyset$ 
and F$_{i + 1}$= F'$_{i}$.

\item[B2)]
If L$_{j}$ is a positive base event, then F$_{i + 1}$= S'$ \cup 
$F'$_{i}$, T$_{i + 1}$= T$_{i}$ and C$_{i + 1 }$= C$_{i } \cup 
${\{}H$_{i}${\}}; where S' is the set of all resolvents of clauses in 
T$_{i}$ with H$_{i}$ on the selected literal L$_{j}$ and []$ \notin $S'. If 
L$_{j}$ is ground then C$_{i + 1 }$= C$_{i}$.

\item[B3)]
If L$_{j}$ is a negative literal, $\neg $L$_{j}$ is not a base event, 
k$>$1 and there is a consistency derivation from ({\{}$ \leftarrow \neg 
$L$_{j}${\}} T$_{i}$ C$_{i})$ to ({\{}{\}} T' C'), then F$_{i + 1 }$= 
{\{}H$_{i}\backslash $L$_{j}${\}} $ \cup $ F'$_{i}$, T$_{i + 1 }$= T' 
and$_{ }$C$_{i + 1}$= C'.

\item[B4)]
If L$_{j}$ is a negative base event, $\neg $L$_{j} \notin $ T$_{i}$ and 
k$>$1, then F$_{i + 1 }$= {\{}H$_{i}\backslash $L$_{j}${\}} $ \cup $ 
F'$_{i}$, T$_{i + 1 }$= T$_{i }$ and $_{ }$C$_{i + 1}$= C$_{i}$.

\item[B5)]
If L$_{j}$ is negative, and there is a constructive derivation from 
({\{}$ \leftarrow \neg $L$_{j}${\}} T$_{i}$ C$_{i})$ to ([] T' C'), then 
F$_{i + 1}$=F'$_{i}$, T$_{i + 1}$=T' and C$_{i + 1}$= C'.

\end{enumerate}
\end{definition}

Rules B1) and B2) are SLDNF resolution steps where A(D) or T act as input 
set, respectively. Rules B3) and B4) allow to continue with the current 
branch by ensuring that the selected literal L$_{j}$ is consistent with 
respect to T$_{i}$ and C$_{i}$. In rule B5) the current branch is dropped if 
there exists a constructive derivation for the negation of the selected 
literal.

Consistency derivations do not rely on the particular order in which 
selection rule R selects literals, since in general, all possible ways in 
which a conjunction $ \leftarrow $L$_{1 } \: \wedge $\ldots $ \wedge \: L_{k}$ can fail should be explored. Each one may lead to a different 
transaction. As a result of this formulation, non-minimal solutions may be 
obtained. They are discarded by means of a simple procedure that rejects those
 that are a superset of the minimal ones.

\begin{example}
\label{exEvents1}
Consider again the same database as in Example \ref{exDedReasoning} and 
assume now that the update request $\iota $Cont(John) is requested. The 
corresponding abductive framework is:

\begin{tabbing}
XXXXX\=XXXXXXXXXX\=\kill
\>EDB $ \cup $ IDB$^{\ast }$: \>Sign(John) \\

\>\>Fail-ex(John) \\

\>\>Cont(x) $ \leftarrow $ Sign(x) $ \wedge \: \neg$Fail-ex(x) \\

\>\>$\iota $Cont(x) $ \leftarrow $Sign(x) $ \wedge  \: \neg \delta 
$Sign(x) $ \wedge \: \delta $Fail-ex(x) \\

\>\>$\iota $Cont(x) $ \leftarrow \iota $Sign(x) $ \wedge \: \neg$Fail-ex(x) 
$ \wedge  \: \neg \iota $Fail-ex(x) \\

\>\>$\iota $Cont(x) $ \leftarrow \iota $Sign(x) $ \wedge \: \delta $Fail-ex(x) \\

\>\>$\delta $Cont(x) $ \leftarrow \delta $Sign(x) $ \wedge \: \neg$
Fail-ex(x) \\

\>\>$\delta $Cont(x) $ \leftarrow $ Cont(x) $ \wedge \: \iota 
$Fail-ex(x) \\

\>\textit{Ab}: \>{\{} $\iota $Sign, $\delta $Sign, $\iota $Fail-ex, 
$\delta $Fail-ex {\}}
\end{tabbing}

\begin{figure}
\centerline{\includegraphics{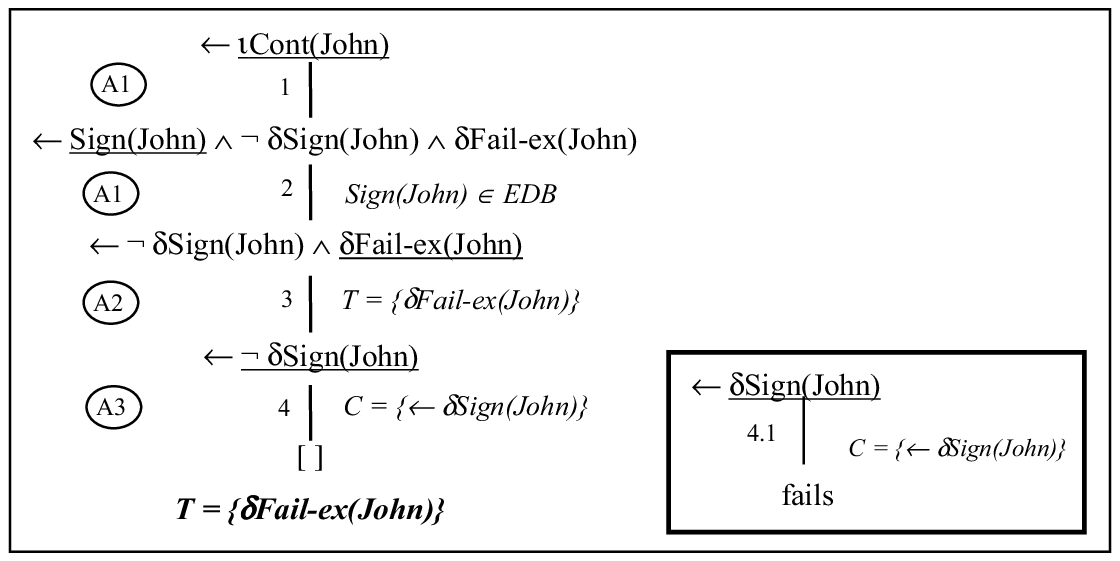}}
\caption{Constructive derivation for $ \leftarrow \iota $Cont(John)}
\label{figEvents1}
\end{figure}

Figure \ref{figEvents1} shows that the abductive interpretation of the event rules of 
Cont provided by the Events Method computes the abductive explanation 
T={\{}$\delta $Fail-ex(John){\}} for $\iota $Cont(John). This is done by 
performing a constructive derivation rooted at $ \leftarrow \iota 
$Cont(John). Circled labels appearing at the left of the derivation are 
references to the rules of the Events Method we have just defined.

The Events Method starts from the update request $ \leftarrow  \iota 
$Cont(John) and uses SLDNF resolution pursuing the empty clause. Steps 1 and 
2 are standard SLDNF resolution steps. At step 3, an 
abducible fact is selected. Then, it is included in the input set T and used 
as input clause if we want to get a successful derivation for $ \leftarrow $ 
$\iota $Cont(John).

Finally, at step 4, a negative base event literal is selected. Then, its 
corresponding subsidiary derivation must be considered, which is shown 
enclosed by the bold box. To ensure failure of this derivation it must be 
guaranteed that $\delta $Sign(John) will not be included into T later on 
during the derivation process. This is achieved by means of an auxiliary set 
C that contains conditions to be satisfied during the whole derivation 
process. These conditions correspond to some of the goals reached in 
subsidiary derivations, as shown at step 4.1 of Figure \ref{figEvents1} 
where the 
condition $ \leftarrow \delta $Sign(John) is included in C. Hence, before 
adding a base event to T we must enforce that it does not falsify any of the 
conditions of C.

Once the empty clause is reached in the primary derivation, the abductive 
procedure finishes and T gives the base events that, applied to the EDB, 
will satisfy the requested update. From this derivation we have that 
T={\{}$\delta $Fail-ex(John){\}}. Then, the request for inserting Cont(John) 
can be achieved by deleting the fact Fail-ex(John).
\end{example}

\begin{example}
Consider now the database of Example \ref{exupdating}. The 
abductive framework relevant to this example is:

\begin{tabbing}
XXXXX\=XXXXXXXXX\=\kill
\>EDB $ \cup $ IDB$^{\ast }$: \>Sign(John) \\

\>\>Fail-ex(John) \\

\>\>Some-cand $ \leftarrow $ Cand(x) \\

\>\>Emp(x) $ \leftarrow $ Cand(x) $ \wedge $ Cont(x) \\

\>\>Cont(x) $ \leftarrow $ Sign(x) $ \wedge \: \neg$Fail-ex(x) \\

\>\>Ic2(x) $ \leftarrow $ App(x) $ \wedge \: \neg$Has-account(x) \\

\>\>Ic4(x) $ \leftarrow $ Cand(x) $ \wedge \: \neg$App (x) \\

\>\>Ic $ \leftarrow $ Ic2(x) \\

\>\>Ic $ \leftarrow $ Ic4(x) \\

\>\>$\iota $Emp(x) $ \leftarrow \iota $Cand(x) $ \wedge \: \iota $Cont(x) \\

\>\>$\iota $Ic $ \leftarrow \iota $Ic2(x) \\

\>\>$\iota $Ic $ \leftarrow \iota $Ic4(x) \\

\>\>$\iota $Ic2(x) $ \leftarrow \iota $App(x) $ \wedge \: \neg$Has-account(x) 
$ \wedge \: \neg \iota $Has-account(x) \\

\>\>$\iota $Ic4(x) $ \leftarrow \iota $Cand(x) $ \wedge \: \neg$App(x) 
$ \wedge \: \neg\iota $App(x) \\

\>\textit{Ab}: \>{\{} $\iota $Cand, $\delta $Cand, $\iota $Cont, 
$\delta $Cont, $\iota $Sign, $\delta $Sign, $\iota $Fail-ex, \\

\>\>$\delta $Fail-ex $\iota $App, $\delta $App, $\iota $Has-account, 
$\delta $Has-account{\}}
\end{tabbing}

\begin{figure}
\centerline{\includegraphics[scale=0.8]{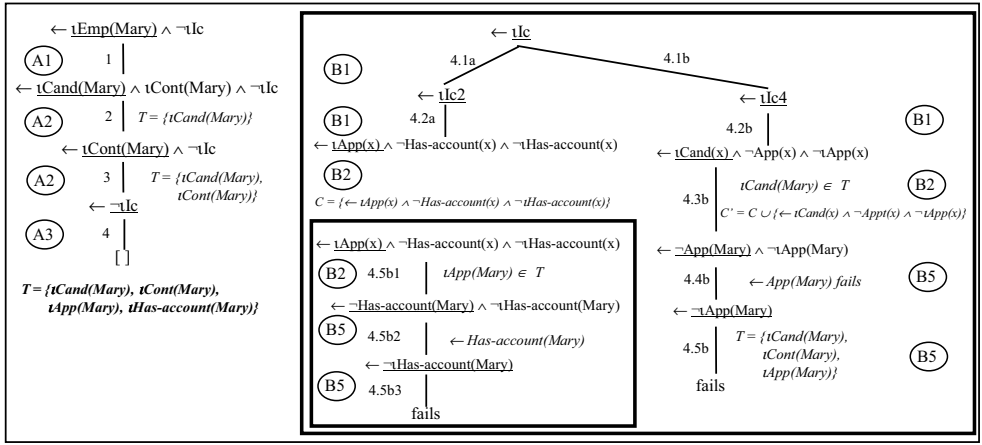}}
\caption{Constructive derivation for $ \leftarrow \iota $ Emp(Mary) 
$ \: \wedge \: \neg \: \iota $Ic}
\label{figEvents2}
\end{figure}

Assume that we want to insert the derived fact Emp(Mary) without violating 
any integrity constraint. In this case, the abductive interpretation of 
$\iota $Emp(Mary) $ \wedge \: \neg \iota $Ic defines the 
possible sets T$_{i}$, {\{}$\iota $Emp(Mary){\}}$ \: \subseteq \: $T$_{i}$, 
that do not induce $\iota $Ic. This is shown in Figure \ref{figEvents2}.

As it happens in the Example \ref{exEvents1}, the Events Method starts from the update request 
$ \leftarrow \iota $Emp(Mary)$\: \wedge \: \neg\iota 
$Ic and uses SLDNF resolution pursuing the empty clause. In this case, base 
event facts $\iota $Cand(Mary) and $\iota $Cont(Mary) are included in the 
translation set T$_{i}$ during steps 2 and 3 of the primary constructive 
derivation. At step 4, the subsidiary consistency derivation (enclosed by 
the bold box) rooted at $\iota $Ic must fail finitely to get the empty 
clause in the constructive derivation. Steps 4.1a and 4.2a of this 
subsidiary derivation are SLDNF resolution steps. After this step, a failed 
goal is reached and it is included in the condition set C to guarantee that 
latter additions to T do not make it succeed. 

Steps 4.1b to 4.4b are SLDNF resolution steps, with the extension that the 
goal is included in C at step 4.3b. At step 4.5b, we have, in turn, a 
subsidiary constructive derivation (not shown in the previous figure) which 
is handled in the same way as the primary constructive one. This derivation 
causes the inclusion of $\iota $App(Mary) into T. Moreover, it must be 
ensured that this inclusion does not violate any of the conditions in C. 
This is done by means of the subsidiary derivation rooted at $ \leftarrow 
\iota $App(x) $ \wedge \: \neg$Has-account(x) $ \wedge \: \neg\iota $
Has-account(x) which, in turn, requires the inclusion of $\iota 
$Has-account(Mary) into T (this is done in the constructive derivation 
associated to step 4.5b3, which is not shown in Figure \ref{figEvents2}). 

Once the empty clause is reached, the abductive procedure finishes and T 
contains the base events that satisfy the requested update. From this 
derivation we have that T$_{i}$={\{}$\iota $Cand(Mary), $\iota $Cont(Mary), 
$\iota $App(Mary), $\iota $Has-account(Mary){\}}. This is the only solution 
that satisfies the requested update in this example.
\end{example}

\subsubsection{Inoue and Sakama's Method}

In this proposal \cite{IS98,IS99}, an abductive logic program ALP is defined as a pair $\prec$
 P, A $\succ$ 
where P is a normal logic program and A is a set of abducible atoms. Given 
an ALP, Inoue and Sakama define a set of production rules, its \textit{transaction program} $\tau $P, 
that declaratively specifies addition and deletion of abductive hypothesis. 
Abductive explanations are then computed by the fixpoint of the transaction 
program using a bottom-up model generation procedure.

They consider two possible kinds of explanations: positive and negative. 
Given an ALP $\prec$ P, A $\succ$ and an observation G, a set of hypothesis E 
is a \textit{positive explanation} for G if P $ \cup $ E $\models$ G and 
P $ \cup $ E is consistent. Similarly, 
a set of hypothesis F is a \textit{negative explanation} for G if 
P$\backslash$F $\models$ G and P$\backslash$F is consistent.

Moreover, they apply abduction also to ``unexplain'' an observation (look 
for antiexplanations). Given a normal logic program P and an observation G, 
a set of hypothesis E is a \textit{positive anti-explanation} for G if P $ \cup $ E $\not \models$ G and E is consistent. 
Similarly, a set of hypothesis F is a \textit{negative anti-explanation} for G if P$\backslash$F $\not \models$ G and E is 
consistent.

Their method has been shown sound and complete for covered acyclic normal 
logic programs. A normal logic program P is covered if, for every rule in P, 
all variables in the body appear in the head. In particular, this 
restriction does not allow having non-ground integrity constraints neither 
existential variables in the body of rules.

There is a clear correspondance between looking for (negative) explanations 
that explain/unexplain an observation and the kind of updates we can 
consider with the event rules. In particular, a positive explanation 
(anti-explanation) E is a set of event facts to be inserted (i.e. a set of 
base insertion events) while a negative explanation (anti-explanation) F is 
a set of event facts to be deleted (i.e. a set of base deletion events). 
Moreover, to explain an observation G that does not hold in the current 
database corresponds to consider the update request $\iota $G, to unexplain 
G (which holds in the current database) corresponds to consider the update 
request $\delta $G.

We may use Inoue and Sakama's method to reason abductively on the event 
rules. We have to:

\begin{enumerate}

\item
Consider the ALP $\prec$ P$^{\ast }$, A $\succ$ 
where P* is the augmented database of P and A is 
the set of base event facts (both insertion and deletion). 

\item
To include the following rules in the transaction program $\tau $P*:

- For each base fact q that belongs to the EDB, we have to include the rule: 
in($\iota $q) $ \to $ false.

- For each base fact q that does not belong to the EDB, we have to include the 
rule: in($\delta $q) $ \to $ false

These rules are needed to distinguish event predicates from database ones 
and to guarantee that an event can be successfully applied. 

Furthermore, since events (and events facts) are handled in this way, we 
also require to apply rule 3 of the definition of transaction program as 
defined in page 346 of \cite{IS99} only to base facts, and not to apply it to 
events facts. This rule states that for any atom $A$ that does not appear in 
the head of any rule in P we must introduce the production rule $out(A) 
\rightarrow \varepsilon$ and if $A$ is not abducible we must introduce also 
$in(A) \rightarrow false$.

\item
To perform abductive reasoning on a derived event fact $\iota $p (resp. 
$\delta $p), we have to \textit{explain} $\iota $p (resp. $\delta $p). On the other hand, to 
perform abductive reasoning on a negative derived event fact $\neg \iota 
$p (resp. $\neg \delta $p), we have to \textit{unexplain} $\neg \iota $p (resp. $\neg 
\delta $p).

\end{enumerate}

As a result of applying Inoue and Sakama's method according to the previous 
transformations we obtain several 
explanations $\prec E_{i},F_{i} \succ$. For each such explanation $\prec E_{i},
F_{i} \succ$, $E_{i}$ corresponds to the abductive explanation that satisfies 
the request while $F_{i}$ contains base events that may not belong to $E_{i}$.

Note that the sets $F_{i}$ play a similar role than the condition set in the 
events method. However, due to the restrictions imposed by Inoue and Sakama's 
method, each $F_{i}$ contains only base event facts while the condition set 
contains more general goals.

Example \ref{exSakama} [adapted from example 3.2 of \cite{IS99}] illustrates 
the use of Inoue and Sakama's method to perform abductive reasoning on the 
event rules.

\begin{example}
\label{exSakama}

\begin{tabbing}
XXXXX\=XXXXXXXXX\=\kill
Let $\prec$ P, A $\succ$ be an ALP where: \\
\>P: \> P $ \leftarrow P \:  \wedge \: \neg$A \\

\>\>Q $ \leftarrow \neg $C \\

\>\>C $ \leftarrow $ \\

\>${\rm A}:$ \>A, C \\

The relevant rules of P* are the following:  \\

\>P: \>$\iota $P $ \leftarrow \iota $Q $ \wedge \: \neg $ A 
$\wedge \: \neg\iota $A \\

\>\>$\iota $Q $ \leftarrow \delta $C \\

\noindent
and the new abducibles atoms are $\delta $C, $\iota $A, $\iota $C, 
$\delta $A.  \\ \\

Then, the subset of $\tau $P* obtained for the previous rules becomes: \\

\>\>in($\iota $P) $ \to $ in($\iota $Q) $ \wedge $ 
out(A)$ \:  \wedge $ out($\iota $A) \\

\>\>in($\iota $Q) $ \to $ in ($\delta $C)
\end{tabbing}

\begin{figure}
\centerline{\includegraphics{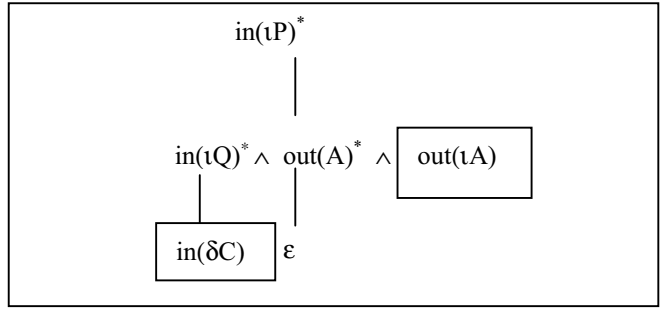}}
\caption{Computing explanations for $ \leftarrow \iota $P}
\label{figSakama}
\end{figure}

Now, to perform abductive reasoning on $\iota $P we have to compute the 
explanations for $\iota $P. Figure \ref{figSakama} shows the behaviour of 
Inoue and Sakama's method in this case. Note that, as a result, we have 
obtained the minimal explanations for $\iota $P: $\prec$E, F$ 
\succ$ = $\prec${\{}in($\delta $C){\}}, {\{}out($\iota $A){\}}$\succ$. 
That means that to insert P we should delete C. 
\end{example}

>From the previous example, we can conclude that Inoue and Sakama's method is 
not only applicable to the update problems they mention in \cite{IS98,IS99}, 
mainly view 
updating and satisfiability checking, but also to the rest of schema 
validation and update processing problems in covered acyclic databases.

Moreover, by explicitly considering the event rules, this method is able to 
perform more precise requests. It is not difficult to see that in \cite{IS99} 
an anti-explanation of P takes two different cases into account: whether P is 
false in the old state of the database and it is not inserted during the 
transition and whether P is deleted during the transition. By considering 
events about P we can be more specific since we can just look for 
anti-explanations of $\iota $P (which corresponds only to the second case). 
A similar claim can be made when looking for explanations. In fact, this 
distinction would allow them also to be able to define and handle dynamic 
integrity constraints. For instance, the dynamic constraint ``it may not be 
inserted Joan as an employee and deleted the Sales department at the same 
time'' may be specified as: $ \leftarrow \iota $Emp(Joan) $ 
\wedge \: \delta $Dept(Sales).

Finally we must note that, in fact, Inoue and Sakama's method would not need 
to use the event rules to obtain the solutions of Example \ref{exSakama}. 
However, we believe that the use of these rules could help this 
method to relax the restrictions it imposes on the programs it deals with. We 
will see in the next section other advantages provided by the event rules when 
more general conditions have to be taken into account.

\subsubsection{The SLDNFA Method }

Denecker and De Schreye propose SLDNFA \cite{DD98}, an extension of SLDNF resolution, to 
deal with abduction in abductive logic programs with negation. Given an 
abductive logic program and a query Q$_{o}$ to be explained, an SLDNFA 
computation can be understood as a process of deriving formulas of the form 
$\forall $ (Q$_{o} \:  \leftarrow \Psi )$, where $\Psi $ is obtained from 
the unsolved goals of the SLDNFA computation.

In fact, SLDNFA distinguishes two kinds of abductive solutions. A \textit{ground abductive solution} for $ 
\leftarrow $Q$_{o}$ is a triple ($\Sigma $',$\Delta $,\textit{$\theta $}) with $\Delta $ a 
finite set of ground abducible atoms and \textit{$\theta $} a substitution of the variables of 
Q$_{o}$, both based on the alphabet $\Sigma $', such that P + $\Delta $ 
$\models \: \forall $ (\textit{$\theta $} (Q$_{o}$ )). Similarly, an \textit{abductive solution} for $ \leftarrow $Q$_{o}$ 
wrt P$^{A}$ is an open formula $\Psi $ containing only equality and 
abducible predicates such that P$^{A} \: \models \: \forall $ (Q$_{o} \:  
\leftarrow \Psi )$ and $\exists (\Psi )$ is satisfiable wrt P$^{A}$. A 
ground abductive solution can be considered as a special case of an 
abductive solution since many ground abductive solutions can be drawn from 
$\Psi $, in general.

SLDNFA has been proved to be sound and complete for failure. Soundness 
ensures that the obtained solutions are correct since they entail the 
initial query and are consistent. Completeness for failure guarantees that 
if there exists a failed SLDNFA-tree for an initial query, then the query 
has no abductive solutions. However, this does not imply that SLDNFA 
generates all ground abductive solutions or all ground solutions satisfying 
some minimality criteria, as already pointed out by Denecker and De Schreye 
\cite{DD98}, page 138. Two variants of SLDNFA, namely SLDNF$^{o}$ and SLDNFA$_{ + 
}$, are defined for which stronger completeness results are proved.

SLDNFA can also be used to perform abductive reasoning on the event rules. 
To do it, we should first rewrite event and transition rules to guarantee 
that they explicitly state the event definition and, thus, that the events 
are correctly handled by SLDNFA. That is, for each event literal $\iota 
$Q(x) appearing in the body of an event or transition rule we should add the 
literal $\neg $Q(x) to that rule, while for each event literal $\delta $Q(x) 
we should add the literal Q(x). This rewriting is equivalent to the addition 
of new rules in Inoue and Sakama's method to guarantee that an event can be 
successfully applied.

Then, we should take the Augmented Database (corresponding to the rewritten 
rules) as the abductive logic program P$^{A}$ where SLDNFA is applied and 
consider base event predicates as the only abducible predicates.

Denecker and De Schreye already mention the applicability of SLDNFA (and its 
variants) to other problems and, in particular, to satisfiability checking. 
>From the use of SLDNFA to perform abductive reasoning on the event rules, we 
can conclude that SLDNFA is also applicable to the rest of schema validation 
and update processing problems. Furthermore, dynamic integrity constraints 
can be easily specified by means of event and transition rules and, thus, 
they could also be handled by SLDNFA.

Example \ref{exDenecker1} [a simplified version of the example in p.119 of 
\cite{DD98}] illustrates the use of SLDNFA to perform abductive reasoning on the 
event rules.

\begin{example}
\label{exDenecker1}

\begin{tabbing}
XXXXX\=XXXXXXXXX\=\kill
Consider the following deductive database: \\

\>\>Q(a) \\

\>\>P(x) $ \leftarrow $ R(x) $ \wedge  \: \neg $Q(x) \\

The abductive logic program P$^{A}$ corresponding to the previous database 
is:  \\

\>$(I.1) $\>$\iota $P(x) $ \leftarrow $R(x) $ \wedge  \: \neg \delta $R(x) 
$ \wedge $ Q(x) $ \wedge \: \neg \delta $Q(x) \\

\>$(I.2) $\>$ \iota $P(x) $ \leftarrow  \: \neg $R(x) $ \wedge \: \iota $R(x) 
$ \wedge \: \neg$Q(x) $ \wedge  \: \neg \iota $Q(x) \\

\>$(I.3) $\>$ \iota $P(x) $ \leftarrow  \: \neg $R(x) $ \wedge \: \iota $R(x) 
$ \wedge $ Q(x) $ \wedge \: \delta $Q(x) \\

\>\textit{Abducibles}: \>{\{} $\iota $R, $\delta $R, $\iota $Q, 
$\delta $Q {\}} \\
\end{tabbing}

\begin{figure}
\centerline{\includegraphics{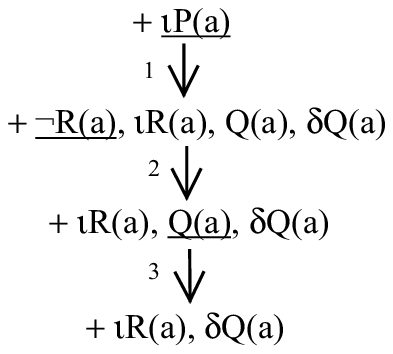}}
\caption{SLDNFA-refutation for $ \leftarrow \iota $P(a)}
\label{figDenecker1}
\end{figure}

An SLDNFA-refutation to find abductive solutions for $\iota $P(a) (which 
corresponds to the initial query $ \leftarrow \iota $P(a)) is shown in 
Figure \ref{figDenecker1}.The ground abductive answer generated by this 
refutation is ($\Sigma $,{\{}$\iota $R(a), $\delta $Q(a){\}},$\varepsilon )$. 
This answer states that the insertion of P(a) may be achieved by the insertion
 of R(a) and the deletion of Q(a).
\end{example}

The use of the event rules provides also other contributions to SLDNFA. For 
instance, it allows SLDNFA to obtain solutions that would not be generated 
otherwise. For instance, SLDNFA (as defined in \cite{DD98}) would not obtain 
any solution to the request insert P(a) in Example \ref{exDenecker1}. The 
reason is that 
SLDNFA considers only positive explanations, but no negative explanations 
(according to the terminology of Inoue and Sakama \cite{IS99}). Therefore, it 
can not generate the deletion of Q(a) which is required to insert P(a). We 
have shown in Figure \ref{figDenecker1} that the use of the event rules allows 
SLDNFA to obtain 
such kind of solutions since abducing events corresponds always to the 
generation of positive explanations although an event may correspond to a 
deletion of a database atom.

Another limitation that is overcome by the use of the event rules is that a 
direct application of SLDNFA to integrity constraint maintenance is not 
incremental. That is, it does not exploit the fact that the old database 
satisfies the integrity constraints, but rechecks blindly all of them \cite{DD98} 
(page 158). Again, such situation can also be overcome if SLDNFA is used in 
conjunction with the event rules, as shown in Example \ref{exDenecker2}.

\begin{example}
\label{exDenecker2}

\begin{tabbing}
XXXXX\=XXXXXXXXX\=\kill
Consider the following deductive database:  \\

\>\>Q(b)   R(b)   Q(c)   R(c)   Q(d)   R(d) \\

\>\>Ic $ \leftarrow $ Q(x) $ \wedge  \: \neg $R(x) \\

The abductive logic program corresponding to the previous database is:  \\

\>$(I.1) $\>$ \iota $Ic $ \leftarrow $Q(x) $ \wedge  \: \neg \delta $Q(x) 
$ \wedge $ R(x) $ \wedge \: \neg \delta $R(x) \\

\>$(I.2) $\>$ \iota $Ic $ \leftarrow  \: \neg $Q(x) $ \wedge \: \iota $Q(x) 
$ \wedge \: \neg$R(x) $ \wedge  \: \neg \iota $R(x) \\

\>$(I.3) $\>$ \iota $Ic $ \leftarrow  \: \neg $Q(x) $ \wedge \: \iota $Q(x) 
$ \wedge $ R(x) $ \wedge \: \delta $R(x) \\

\>\textit{Abducibles}: \>{\{} $\iota $Q, $\delta $Q, $\iota $R, $\delta $R {\}} \\
\end{tabbing}

\begin{figure}
\centerline{\includegraphics{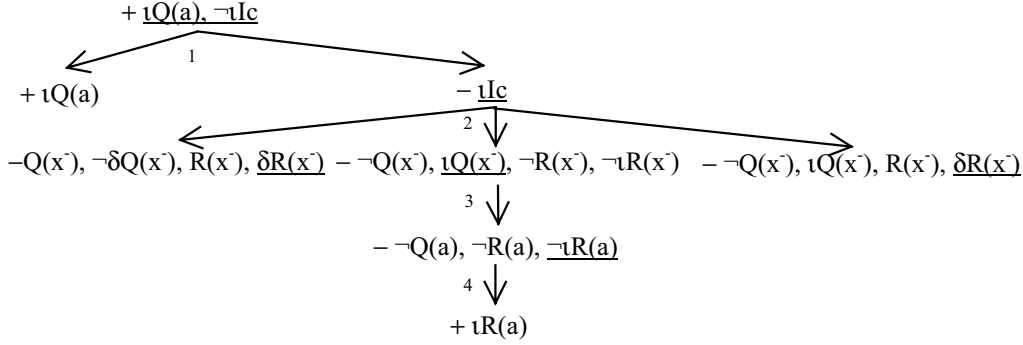}}
\caption{SLDNFA-refutation for $ \leftarrow \iota $Q(a)$\: \wedge  \: \neg 
\iota $Ic}
\label{figDenecker2}
\end{figure}

An SLDNFA-refutation to find abductive solutions for $\iota $Q(a) that do 
not violate integrity constraints (which corresponds to the initial query $ 
\leftarrow \iota $Q(a)$ \wedge  \: \neg \iota $Ic) is shown in 
Figure \ref{figDenecker2}.

The ground abductive answer generated by this refutation is ($\Sigma 
$,{\{}$\iota $Q(a), $\iota $R(a){\}},$\varepsilon )$. Note that, in this 
case, integrity constraint maintenance has taken into account that integrity 
constraints are satisfied before the update since it has only considered the 
events that could induce a violation of Ic (and not all database facts 
involved in the definition of Ic, as SLDNFA would do in the absence of the 
event rules).
\end{example}

\subsubsection{How many Methods do we really need to deal with Database 
Problems?}

In general, research related to database problems is still looking for 
methods able to deal only with specific problems. However, we have shown that 
all problems are either of abductive or deductive nature and, thus, they can 
be formulated in terms of just two forms of reasoning. Therefore, we may 
conclude from our results that at most two different procedures are enough to 
handle all of them.

In fact, it could also be argued that just one single general procedure 
would be enough since it has been shown that either abduction as well as 
deduction can be realized in terms of the other. For instance, \cite{Bry90,BEST98} have proposed a procedure that allows to perform abductive reasoning 
by means of deduction, while \cite{DD98,IS99} have shown that their abductive 
procedures can also be used for deduction. Clearly, this strengthens our 
claim that we do not need a different method to deal with each schema 
validation and update processing problem. 

The discussion about whether it would be better to have just one single 
method or two is far beyond the scope of this paper since it requires to be 
further investigated. From our intuition, we think that it will be difficult 
to define a single method which is as efficient to perform abductive 
reasoning as it is to perform deductive reasoning since, as we have shown in 
the paper, the nature of the corresponding problems is intrinsically 
abductive or deductive. In this sense, we believe that it is difficult to 
incorporate the optimizations required to efficiently deal with deductive 
problems into a method based on abductive reasoning (and the other way 
around). However, this is still an open problem and it is a challenger 
research.

\section{Using the Event Rules to Perform General Abductive Reasoning}

We have shown how the event rules can be used to deal with several schema 
validation and update processing problems and we have illustrated how 
several deductive and abductive procedures can be used to reason on them. In 
this section, we sketch how the event rules can be used also to solve 
general abductive problems in addition to the database problems considered 
before. Obviously, we must take into account that whenever we use the event 
rules we obtain solutions that minimize the difference between the old and 
the new sates and that this is not necessarily a requirement that the 
abductive explanations must satisfy in general.

We illustrate, by means of the following example, how could we use the event 
rules to deal with a typical abductive task like fault diagnosis and how the 
expected solutions to this problem can be abduced from these rules.

\begin{example}
\label{exabducgeneral}

The relevant part of the abductive framework of 
this example is the following: 

\begin{tabbing}
XXXXXXXXX\=XXXXXXXXXXXXXXXXXXXXXXXXX\=\kill
EDB $ \cup $ IDB$^{\ast }$: \>Lamp(L1) \\
\>Battery(C1,B1) \\

\>Faulty-lamp $ \leftarrow $ Lamp(x) $ \wedge $ Broken(x) \\
\>Backup(x) $ \leftarrow $ Battery(x,y) $ \wedge  \: \neg $Unloaded(y) \\

\>Faulty-lamp $ \leftarrow $ Power-failure(x) $ \wedge  \: \neg $Backup(x) \\
\>Unloaded(x) $ \leftarrow $ Dry-cell(x) \\

\>$\iota $Faulty-lamp $ \leftarrow $Faulty-lamp$^{n} \:  \wedge \: \neg$Faulty-lamp \\

\>Faulty-lamp$^{n} \:  \leftarrow  \: \iota $Power-failure(x) $ \wedge $ 
Backup(x) $ \wedge  \: \delta $Backup(x) \\

\>Faulty-lamp$^{n} \:  \leftarrow  \: \iota $Power-failure(x) $ \wedge \: \neg$Backup(x) $ \wedge  \: \neg \iota $Backup(x) \\

\>Faulty-lamp$^{n} \:  \leftarrow $ Lamp(x) $ \wedge \: \neg \delta $Lamp(x) 
$ \wedge  \: \iota $Broken(x) \\

\>$\delta $Backup(x) $ \leftarrow $Battery(x,y) $ \wedge \: \iota $Unloaded(y)
 $ \wedge $ Backup$^{n}$(x) \\

\>$\iota $Unloaded(x) $ \leftarrow \: \iota $Dry-cell(x) \\

\textit{Ab}:    {\{} $\iota $Broken, $\delta $Broken, $\iota $Power-failure, 
$\delta $Power-failure, $\iota $Dry-cell, $\delta $Dry-cell {\}} \\
\end{tabbing}

Within this framework, we can use the event rules to 
detect that a faulty lamp problem is caused by a broken lamp or by a power 
failure of a circuit without backup, that is, a loaded battery. To do it, we 
have selected the Events Method among the three possible abductive 
procedures considered in Section \ref{sectAbdProd}.

\begin{figure}
\centerline{\includegraphics{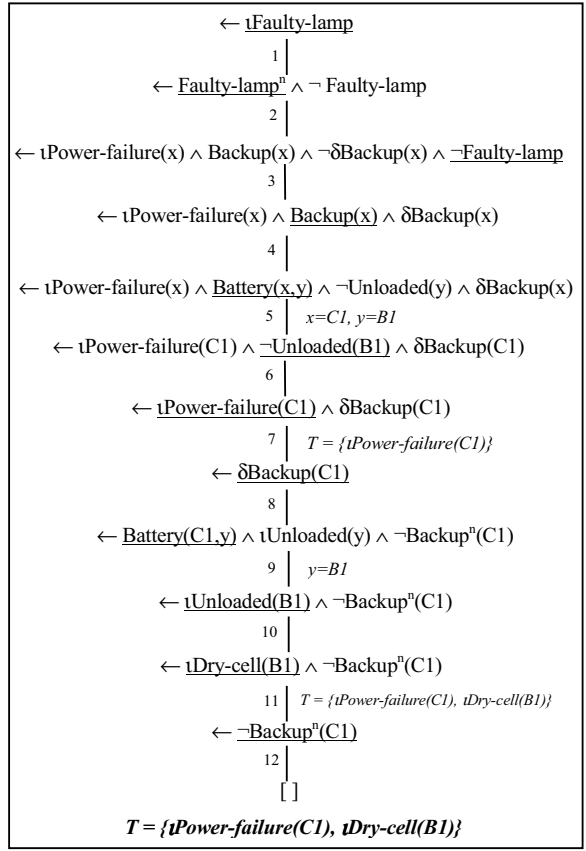}}
\caption{Constructive derivation for $ \leftarrow \iota $Faulty-lamp}
\label{figSec7}
\end{figure}

Figure \ref{figSec7} shows how, given the goal $ \leftarrow  \: \iota 
$Faulty-lamp and EDB $ \cup $ IDB$^{\ast }$, the Events Method obtains the 
abductive solution T={\{}$\iota $Power-failure(C1), $\iota 
$Dry-cell(B1) {\}}, which is a possible solution for having a faulty lamp.

The Events Method would also obtain the solution T={\{}$\iota 
$Broken(L1){\}} by considering the rule Faulty-lamp$^{n}$(x) $ \leftarrow $ 
Lamp(x) $ \wedge \: \delta $Lamp(x) $ \wedge  \: \iota $Broken(x) at 
step 2 of Figure \ref{figSec7}, and other possible solutions, like for 
instance T={\{}$\iota $Power-failure(C2){\}}, by considering the 
rule Faulty-lamp$^{n}$(x) $ \leftarrow  \: \iota $Power-failure(x) $ \wedge$ 
Backup(x) $ \wedge  \: \neg \iota $Backup(x) also at step 2. The 
resulting derivations are not shown in the tree above.
\end{example}

\section{Conclusions and Further Work}

We have shown that database schema validation and update processing problems 
can be classified into problems of either deductive or abductive nature 
according to the reasoning paradigm that is more adequate to solve them. 
This has been done by making explicit the exact changes that occur in a 
transition between two consecutive states of the database by means of the 
event rules \cite{Oli91} and by performing deductive and abductive reasoning on 
these rules. In this way, we have distinguished between\textit{ deductive problems}, concerned with 
computing the changes on derived predicates induced by a transaction, and 
\textit{abductive problems}, concerned with determining the possible transactions that satisfy a set of 
changes on derived predicates. We have also shown that deductive and 
abductive problems can be combined, thus defining more complex update 
problems.

Thus, problems like materialized view maintenance, integrity constraint 
checking or condition monitoring are considered as naturally deductive, 
while problems like view updating, integrity constraint maintenance or 
enforcing condition activation as naturally abductive.

By taking only a unique set of rules and two forms of reasoning into account 
to specify and deal with all these problems, we have shown that it is 
possible to provide general methods able to uniformly deal with several 
database updating problems at the same time. This suggests that future 
research in this field should be aimed at providing general methods instead 
of proposing specific methods for solving a particular problem like has been 
traditionally done in the past. Moreover, all these problems could be 
uniformly integrated into a database update processing system.

We have also shown how some existing general deductive and abductive 
procedures may be used to reason on the event rules. In this way, we have 
shown that these procedures can be used to deal with all database problems 
considered in this paper. This has been illustrated by means of examples and 
some additional benefits gained by these procedures when reasoning on the 
event rules have also been pointed out. Moreover, we have sketched how the 
event rules could be used to solve general abductive problems in addition to 
database schema validation and update processing problems. 

The results presented in this paper may be extended at least in three 
different directions. First, our framework could be generalized to databases 
that allow recursive rules. Second, to deepen in the study of the 
application of the event rules to current abductive procedures to provide an 
efficient implementation of abductive reasoning on the event rules. Third, the 
advantages and inconveniencies of using the event rules to perform general 
abductive reasoning should be further investigated.

\textbf{Acknowledgements}

This work has been partially supported by the CICYT PRONTIC program project 
TIC97-1157.

\bibliography{Tplpref}

\end{document}